\documentclass[12pt]{iopart}

\usepackage{iopams}
\usepackage{cite}
\usepackage{graphicx}
\usepackage{amssymb}
\usepackage{textcomp}
\usepackage{multirow}
\usepackage{booktabs}
\usepackage[british]{babel}
\usepackage{color}
\usepackage[T1]{fontenc}
\usepackage{cases}
\usepackage{mathrsfs}

\newcommand{\eqref}[1]{(\ref{#1})}

\begin{document}

\title[Continuous Time Random Walk in a velocity field]
{Continuous Time Random Walk in a velocity field: Role of domain growth,
Galilei-invariant advection-diffusion, and kinetics of particle mixing}

\author{F. Le Vot$^{\dagger}$, E. Abad$^{\ddagger}$, R. Metzler$^{\sharp}$,
and S. B. Yuste$^{\dagger}$}
\address{$\dagger$ Departamento de F\'{i}sica and Instituto de Computaci\'on Cient\'{i}fica Avanzada, Universidad de Extremadura, 06071 Badajoz, Spain\\
$\ddagger$ Departamento de F\'{i}sica Aplicada and Instituto de Computaci\'on Cient\'{i}fica Avanzada, Centro Universitario de M\'erida, Universidad de Extremadura, 06800 M\'erida, Spain\\
$\sharp$ Institute for Physics \& Astronomy, University of Potsdam,
Karl-Liebknecht-Str 24/25, 14476 Potsdam, Germany}

\begin{abstract}
We consider the emerging dynamics of a separable Continuous Time Random Walk (CTRW)
in the case when the random walker is biased by a velocity field in a uniformly
growing domain. Concrete examples for such domains include growing biological
cells or lipid vesicles, biofilms and tissues, but also macroscopic systems
such as expanding aquifers during rainy periods, or the expanding universe.
The CTRW in this study can be subdiffusive, normal diffusive or superdiffusive,
including the particular case of a L\'evy flight. We first consider the case when
the velocity field is absent. In the subdiffusive case, we reveal an interesting
time dependence of the kurtosis of the particle probability density function. In
particular, for a suitable parameter choice, we find that the propagator, which is fat tailed
at short times, may cross over to a Gaussian-like propagator. We subsequently incorporate
the effect of the velocity field and derive a bi-fractional diffusion-advection
equation encoding the time evolution of the particle distribution. We apply this
equation to study the mixing kinetics of two diffusing pulses, whose peaks move
towards each other under the action of velocity fields acting in opposite
directions. This deterministic motion of the peaks, together with the diffusive
spreading of each pulse, tends to increase particle mixing, thereby counteracting
the peak separation induced by the domain growth. As a result of this competition,
different regimes of mixing arise. In the case of L\'evy flights, apart from the
non-mixing regime, one has two different mixing regimes in the long-time limit,
depending on the exact parameter choice: In one of these regimes, mixing is mainly
driven by diffusive spreading, while in the other mixing is controlled by the
velocity fields acting on each pulse. Possible implications for encounter-controlled
reactions in real systems are discussed.
\end{abstract}

\section{Introduction}

Since its introduction by Montroll and Weiss in the mid 60s \cite{montroll},
the celebrated Continuous Time Random Walk (CTRW) model has found widespread
application in statistical physics and beyond, notably in the study of problems
as diverse as charge carrier transport in disordered media such as amorphous
semiconductors \cite{scher,scher1,henning}, luminescence quenching \cite{barzykin},
morphogen gradient formation \cite{hornung,yuste}, the diffusive motion of
water molecules in the hydration shell around proteins \cite{ling}, the
relative motion of monomers in a protein molecule \cite{smith}, the motion of
protein channels in the cell membrane \cite{diego} of lipid and insulin
granules \cite{stas,lene}, or of active transport \cite{granick,jae} in living
biological cells, up to chemical tracer dispersion in groundwater aquifers
\cite{edely,grl}, and even ageing effects in stock markets \cite{scalas,masoliver}.
Several review articles and monographs have devoted substantial parts on CTRW
\cite{pt-scher,pt-yossi,nature-mike,bouchaud,report,report1,igorsm,pccp,pt1,pt2,
hughes}. Specific properties of the CTRW concern the concept of weak ergodicity
breaking \cite{bouchaud_web,eli_rebenstok,eli_bel,irwin,stas_pnas} and ageing
\cite{bouchaud_age,eli_age,johannes,johannes1,manzo}.

In the original version of the CTRW model, the probability density functions
(PDFs) of the waiting times (also called trapping or sojourn times) between
successive jumps and of the lengths of individual jumps are assumed to
decouple, that is, are independent of each other. In the case when the PDF
of the waiting times $\tau$ is fat-tailed and scale-free, $\psi(\tau)\simeq
\tau^{-1-\alpha}$ with $0<\alpha<1$, and the variance of the jump length PDF
is finite, the long-time limit of the model is known to yield anomalous diffusion
in the subdiffusive range with the mean-squared displacement $\langle x^2(t)\rangle
\simeq t^{\alpha}$ \cite{montroll,scher,scher1,bouchaud,report,report1,igorsm,pccp}.
In the more general case, given specific forms of the waiting time and jump length
PDFs, the emerging behaviour may be subdiffusive, diffusive, or superdiffusive.
This versatility of the model, together with the fact that it can be shown to be
equivalent to a generalised master equation \cite{yossi_prl} and a fractional
diffusion equation (FDE) in the anomalous diffusion case \cite{compte,mebakla,
mebakla1} makes the CTRW a popular choice to model anomalous transport. One of
the advantages of the FDE formulation is that it can incorporate the effect of
external force fields and various boundary conditions in a natural and transparent
way \cite{report,report1,meklabvp}. Further generalisations are also possible,
for instance, accounting for the effect of finite lifetimes of tracer particles
(so called ``evanescent'' or ``mortal walkers'') \cite{katja,yuste, mortal, evanescent},
or of chemical reactions occurring at random times and locations \cite{yossi_ws,henry,seki,igor_francesc,henry1,
yadav,fedotov,yadav1,froemberg,gafi, MendezBook}.

Recently, it has been discussed how separable CTRWs need to be formulated when the domain,
on which the process is running off, is itself explicitly evolving in time.
 In particular, it has been shown that an FDE can be derived \cite{levot,angst, levot1, abad}, whence
 the case of normal diffusion on an evolving domain \cite{Yuste2016} is recovered in the
 appropriate limit.  The obtained FDE applies when the diffusing particles
stick to the evolving domain, implying that they experience a drift even when they do not jump.
The interplay between diffusive transport and the drift associated
with the growth or contraction of the embedding medium gives rise to the onset of striking effects. These
include an enhanced memory of the initial condition \cite{Yuste2016, levot} and the slowing-down and even the premature halt of encounter-controlled reactions \cite{VEAY2018, EYAV2018, AEVY2019}. In the case of subdiffusive particles evolving on an exponentially shrinking domain, a so-called Big Crunch may happen. This phenomenon was first discussed in reference \cite{levot}; it consists in the collapse of an initial particle distribution with finite extent to a delta function as a result of the strong localisation caused by the domain contraction.

Concrete examples of expanding (or shrinking) domains include biological cells
in interphase \cite{alberts}, growing biofilms \cite{biofilm,biofilm1}, growing
biological tissues \cite{tissue,tissue1}, and growing or shrinking lipid vesicles
\cite{vesicle,vesicle1}. The latter can be controlled easily, for instance, by
adjusting osmotic pressure in solution \cite{vesicle2}. Water drops, puddles,
or aqueous solutions in a Petri dish shrink simply by evaporation \cite{carrier}.
On a geophysical scale, groundwater aquifers may be recharged by major flood
events and thus the volume for tracer dispersion increased \cite{yang,zhang}.
On Earth, the subducted oceanic lithosphere is stretched by the convective mantle,
and both are homogenized, among other mechanisms, by diffusion \cite{allegre,kellogg}.
Finally, expanding domains are traditionally considered in cosmological
models describing the diffusion of cosmic rays in the expanding universe
\cite{berezinsky,batista}.

In the present paper, we consider the case where both a domain growth process
and a left-right bias of the random walk are simultaneously at play. In reference
\cite{levot1}, such a combination was considered for the case where the bias
stems from a force field that only manifests itself at the time of each jump.
To model the effect of the force field, a non-symmetric jump length distribution
was used, and the corresponding Fokker-Planck equation (FPE) was obtained.
This approach has been recently used to deal with the Ornstein-Uhlenbeck
process on a growing domain \cite{LVYA2019}.

In contrast, here we will focus on the case of a bias arising from a velocity field.
This field is still at play while the walker rests between jumps. Direct realisations
of such a situation could occur in biological cells in the presence of active,
motor-driven motion \cite{jaehyung}, in suspended giant vesicles simply by
gravity. In subsurface aquifers the flow field corresponds to groundwater
streams towards a spring or well.

In a static domain, a constant force field and a constant velocity field
yield the same type of advection-diffusion equation as long as the particles
are normal-diffusive. However, this is no longer true when the CTRW becomes
subdiffusive \cite{chen}. Then, a constant force field is assumed to
act only on particles when they are not trapped, while in a constant velocity field the particle is
constantly advected. The former case may, for instance, correspond to charge
carriers in amorphous semiconductors (in which they are trapped at impurities)
in the presence of an electrical field
\cite{scher} while the latter may correspond to a particle moving in a flowing
complex environment such as an actin gel. Of course, this lack of equivalence
carries over to the case of an evolving domain. In \cite{Cairoli2018}
a fractional diffusion-advection equation (FDAE) was derived for a subdiffusive
CTRW in the presence of a constant velocity field. One of our main goals will
be to extend their result by considering a CTRW which takes place in a uniformly
growing domain, and also including the case of superdiffusion.

In our derivation of the sought FDAE, we will first study the case in absence of
the velocity field, to see that the domain growth itself may induce interesting
behaviour of the moments of the particle position. For instance, in the
subdiffusive case, when the physical domain grows with a power-law rate we see
that a fat-tailed propagator may evolve into a Gaussian-like propagator for
a suitable parameter choice.

After deriving the FDAE we will use it to study the mixing kinetics of a pair
of diffusive pulses evolving on a growing domain in the presence of the velocity field,
and we will discuss possible implications of the results for the kinetics of encounter-controlled
reactions in real systems \cite{VEAY2018, EYAV2018, AEVY2019}.

The remainder of this paper is organised as follows. In section \ref{sec:unbiased}
we first recall the main results for a symmetric CTRW on a static domain, and we
subsequently discuss how the kurtosis of a subdiffusive walk changes when the
initial domain grows uniformly in time. In section \ref{sec:biased} we carry out
a similar programme for a CTRW subject to the action of a velocity field, and we
will derive the relevant FPE for the case of a uniformly growing domain. In section
\ref{sec:pulses} we study the mixing of diffusive pulses that are biased by velocity
fields acting in opposite directions. Finally, in section \ref{sec:summary}
we summarise our main results, discuss their possible relevance for
encounter-controlled reactions in real systems, and outline possible extensions
of the present work.

\section{CTRW in the absence of the velocity field: Static versus growing domain}
\label{sec:unbiased}

In this section we compare the behaviour of a separable, symmetric CTRW evolving
on a one-dimensional static domain with the same walk on a \emph{uniformly growing\/}
domain. For both cases we discuss the difference in the behaviour of the moments of
the particle position. In the case of the growing domain special emphasis is paid
on the interesting onset of a time dependence at the level of the kurtosis. This
section is also intended to introduce the general concepts and thus prepare the
reader for the case of a CTRW subject to a constant velocity field, discussed
in section \ref{sec:biased}.

\subsection{Static domain}
\label{subsec:statdom}

We start with a brief reminder of the derivation of a bifractional equation
describing the diffusive limit of a fat-tailed CTRW. For further details
we refer to the review \cite{report} and to the recent monograph
\cite{yossi_ws}.

Consider a particle performing a one-dimensional, symmetric CTRW with decoupled
jump length and waiting time PDFs, respectively denoted by $\lambda(y)$ and
$\varphi(t)$. Since the random walk is symmetric, the jump length PDF reflects
this symmetry, $\lambda(y)=\lambda(-y)$. The Fourier-Laplace transform
$W_0(k,u)$ of the particle's position PDF $W_0(x,t)$ is known to obey the
Montroll-Weiss relation \cite{montroll,scher,scher1}
\begin{equation}
\label{MW0eq}
W_0(k,u)=\frac{\Phi(u)}{1-\varphi(u)\lambda(k)}W_0(k,0),
\end{equation}
where $W_0(k,0)=W_0(k,t=0)$ denotes the (Fourier-transformed) initial condition,
$\varphi(u)$ is the Laplace transform of the waiting time PDF, and $\Phi(u)=u^{
-1}(1-\varphi(u))$ is the Laplace transform of the sticking probability $\Phi(t)
=1-\int_0^tdt'\varphi(t')$ for not performing a jump up to time $t$. Here, we have
used standard definitions of the Laplace transform
\begin{equation}
\label{LTdefi}
\mathscr{L}[f(t)]=f(u)=\int_0^\infty e^{-u t} f(t)\, dt
\end{equation}
and of the Fourier transform
\begin{equation}
\label{FTdefi}
\mathscr{F}[f(y)]=f(k)=\int_0^\infty e^{-i k y} f(y)\, dy.
\end{equation}
The subscript ``0'' in the definition of $W_0$ indicates that this quantity
refers to the case where the velocity field is absent.

The diffusive limit of the CTRW process and the associated FDE are obtained
from the long-time behaviour of $\varphi(t)$ and from the large-$|y|$ behaviour
of $\lambda(y)$, which respectively correspond to the small-$u$ behaviour
of $\varphi(u)$ and to the small-$k$ behaviour of $\lambda(k)$. For these
transforms of the respective PDFs we use the well-known forms \cite{hughes}
\begin{equation}
\label{varphiusmall}
\varphi(u)= 1-(\tau u)^\alpha+\ldots
\end{equation}
with $0<\alpha\le 1$, and
\begin{equation}
\label{laksmall}
\lambda(k)= 1-\sigma^\mu\,|k|^\mu+\ldots
\end{equation}
with $0<\mu\le 2$. This implies the asymptotic power-law forms $\varphi(t)
\sim\tau^{\alpha}/t^{1+\alpha}$ for $t\gg\tau$ with $0<\alpha<1$ as well as
$\lambda(y)\sim\sigma^{-\mu}|y|^{-1-\mu}$ for $|y|\gg\sigma$ with $0<\mu<2$.
In the limit $\alpha=1$ the characteristic waiting time $\langle t\rangle=
\tau$ is finite, and typical choices are either $\varphi(t)=\delta(t-\tau)$
or $\varphi(t)=\tau^{-1}\exp(-t/\tau)$. Similarly, in the limit $\mu=2$ the
variance $\langle y^2\rangle=\sigma^2$ of the jump lengths is finite, and
the typical choice is the Gaussian form $\lambda(y)=(2\pi\sigma^2)^{-1/2}
\exp(-y^2/[2\sigma^2])$ \cite{hughes}. In particular, the choice $\alpha=1$
and $\mu=2$ then yields Brownian diffusion with finite characteristic waiting
time and jump length variance, and thus a Gaussian position PDF. In contrast,
$0<\alpha<1$ and $\mu=2$ lead to (fractional) subdiffusion, and $\alpha=1$
and $0<\mu<2$ correspond to (fractional) superdiffusion (L\'{e}vy flights)
(for more details, see \cite{hughes,report}).

Inserting equations \eqref{varphiusmall} and \eqref{laksmall} into \eqref{MW0eq}
yields
\begin{equation}
\label{W0kuljr}
W_0(k,u)=\frac{W(k,0)}{u+K^\mu_\alpha|k|^\mu u^{1-\alpha}},
\end{equation}
or, equivalently,
\begin{equation}
\label{W0kuljrbis}
uW_0(k,u)-W(k,0)=-K_\alpha^\mu|k|^\mu u^{1-\alpha} W_0(k,u).
\end{equation}
This is a diffusion equation in Fourier-Laplace space. In direct space, the equation reads
\begin{equation}
\label{W0Eq}
\frac{\partial}{\partial t} W_0(y,t)=K_\alpha^\mu\nabla^\mu_y\,{_0}\mathcal{D}_t^{
1-\alpha}W_0(y,t),
\end{equation}
where $K_\alpha^\mu=\sigma^\mu/\tau^\alpha$ is the anomalous diffusion coefficient
of dimension $\mathrm{cm}^{\mu}/\mathrm{sec}^{\alpha}$, $_0\mathcal{D}_t^{1-\alpha}$
stands for the Gr{\"u}nwald-Letnikov (GL) fractional derivative, and $\nabla^\mu_y$ is
the Riesz fractional operator \cite{report}. The latter is defined via the relation
$\mathscr{F} \nabla^\mu_y f(y)=-|k|^\mu f(k)$ in Fourier space.

The GL fractional operator has the property \cite{Podlubny1999}
\begin{equation}
\label{GLLaplace}
\mathscr{L}{_0}\mathcal{D}_t^{1-\alpha}f(t)=u^{1-\alpha}f(u).
\end{equation}
This fractional operator is equivalent to the Riemann-Liouville (RL) fractional
derivative
\begin{equation}
\label{GLeqRL}
_0\mathcal{D}_t^{1-\alpha}f(t)={_{0}}D_t^{1-\alpha}f(t)
\end{equation}
provided that both operators are applied to sufficiently smooth functions $f(t)$ at
$t=0$, that is, when the condition \cite{Podlubny1999}
\begin{equation}
\label{GLeqRLcond}
\lim_{t\to 0} \int_0^t (t-\tau)^{\alpha-1}\, f(\tau)\, d\tau \to 0
\end{equation}
holds. Unless otherwise specified, we will henceforth assume that this is the case
and therefore use the RL fractional derivative in what follows. The RL operator
$_0D_t^{1-\alpha}$ is simply the first derivative of the RL fractional integral
\begin{equation}
\label{RLIntegral}
_0D_t^{-\alpha}=\frac{1}{\Gamma(\alpha)}\int_0^td\tau\frac{f(\tau)}{(t-\tau)^{
1-\alpha}}.
\end{equation}

For the special case $\mu=2$ implying a finite variance of the jump-length PDF, it
is possible to obtain the behaviour of the moments associated with the solution
$W_0(y,t)$ of equation \eqref{W0Eq} either from the exact solution or from the
corresponding hierarchy of differential equations. For $0<\mu<2$, in contrast,
only fractional moments of order $\nu$ exist as long as $0<\nu<\mu$. For a
particle initially located at the origin, $W_0(y,0)=\delta(y)$, the well-known
propagator for $\mu=2$ reads
\begin{equation}
\label{solH}
W_0(y,t)=\frac{1}{\sqrt{4K_\alpha t^\alpha}}H_{1,1}^{1,0}\left[\left.\frac{|y|}{
\sqrt{4K_\alpha t^\alpha}}\right|\begin{array}{ll}(1-\alpha/2,\alpha/2)\\(0,1)
\end{array}\right],
\end{equation}
where $K_\alpha\equiv K_\alpha^2$. In result \eqref{solH}, $H_{1,1}^{1,0}[\cdot]$
stands for a Fox $H$-function \cite{mathai}. For $0<\alpha<1$, the above propagator
displays a non-differentiable peak (cusp) at the origin. The solution \eqref{solH}
is equivalent to the series representation
\begin{equation}
W_0(y,t)=\frac{1}{\sqrt{4 K_\alpha t^\alpha}}\sum_0^\infty \frac{(-1)^n}{n!
\Gamma(1-\alpha[n+1]/2)}\left(\frac{x^2}{K_\alpha t^\alpha}\right)^{n/2}.
\end{equation}
Employing standard theorems for the Fox functions \cite{mathai} one can show that
for $|y|\gg\sqrt{K_\alpha t^\alpha}$ the following asymptotic stretched Gaussian
behaviour emerges \cite {report}:
\begin{eqnarray}
\label{stretchedG}
W(y,t)&\sim&\frac{1}{\sqrt{4\pi K_\alpha t^\alpha}}\sqrt{\frac{1}{2-\alpha}}
\left(\frac{2}{\alpha}\right)^{(1-\alpha)/(2-\alpha)}
\left(\frac{|y|}{\sqrt{K_\alpha t^\alpha}}\right)^{-(1-\alpha)(2-\alpha)}\nonumber\\
&&\times\exp\left(-\frac{2-\alpha}{2}\left(\frac{\alpha}{2}\right)^{\alpha/(2-\alpha)}
\left[\frac{|y|}{\sqrt{K_\alpha t^\alpha}}\right]\right).
\end{eqnarray}
From equation \eqref{solH} (or from symmetry arguments) it is immediately clear
that odd moments vanish. In turn, the behaviour of the even moments is strongly
influenced by the stretched Gaussian behaviour \eqref{stretchedG}. One finds \cite{report}
\begin{equation}
\label{all-moments}
\langle y^{2n}(t)\rangle_0=(2n)!\frac{(K_\alpha t^{\alpha})^n}{\Gamma(1+n\alpha)},
\quad n=0,1,\ldots
\end{equation}
Higher order integer moments can be expressed in terms of the variance $\langle y^2
\rangle$ as follows,
\begin{equation}
\label{2nth-moment}
\langle y^{2n}\rangle_0=\frac{(2n)![\Gamma(1+\alpha)]^n}{2^n\Gamma(1+n\alpha)}
\langle y^2\rangle_0^n.
\end{equation}
For $\alpha=1$ (normal diffusion) one recovers the moments characterising
the typical Gaussian propagator,
\begin{equation}
\langle y^{2n}\rangle_0=\frac{(2n)!}{2^n n!}\langle y^2\rangle_0^n=(2n-1)!!
\langle y^2\rangle_0^n.
\end{equation}
In particular, equation \eqref{2nth-moment} can be used to calculate the kurtosis.
This quantity is a measure of the ``tailedness'' of a given probability distribution,
defined as
\begin{equation}
\label{kurtosis}
\beta_2=\frac{\langle(y-\langle y\rangle)^4\rangle}{\langle(y-\langle y\rangle)^2
\rangle^2}.
\end{equation}
Recall that for a normal distribution one has $\beta_2=3$ in one dimension. In
contrast, a fat-tailed distribution exhibits a large skewness or kurtosis, relative
to that of a normal distribution. Distributions with $\beta_2>3$ are called leptokurtic
(as opposed to distributions with $\beta_2<3$, which are termed platykurtic). For the
particular case $\mu=2$ described by result \eqref{all-moments} one finds
\begin{equation}
\beta_2=\frac{\langle y \rangle_0^4}{\langle y^2\rangle_0^2}=6\frac{[\Gamma(
1+\alpha)]^2}{\Gamma(1+2\alpha)}=\frac{3\Gamma(\alpha)\Gamma(1+\alpha)}{\Gamma(
2\alpha)}.
\label{Beta2_Static}
\end{equation}
Thus, $\beta_2$ decreases monotonically from $\beta_2=6$ for $\alpha=0$ to $\beta
_2=3$ for $\alpha=1$ as $\alpha$ increases. In other words, the tails become less
fat with increasing $\alpha$, and in this sense the distribution becomes more
Gaussian-like. However, for any value $\alpha<1$, the distribution remains
leptokurtic with $\beta_2>3$.

\subsection{Growing domain}

Next we compare the behaviour of positive integer moments up to the kurtosis
of the particle distribution for a symmetric CTRW with their counterparts in
a uniformly growing domain.

In the growing domain, the coordinate $y$ of a physical point (hereafter
also termed ``physical coordinate'' or ``Eulerian coordinate'') is no longer
stationary, since each physical point is advected by the growing domain. Thus,
the distance between a physical point at $y$ and the origin $0$ changes in
time as the domain expands. In the following, we will also assume that a random
walker ``sticks'' to the physical medium while it does not jump \footnote{For instance,
a tracer bead intermittently stuck in an expanding hydrogel.}---consequently
the walker is also advected by the medium as it expands. It is convenient to
describe the time evolution of $y$ in terms of its initial position $x\equiv
y_0$, hereafter termed ``Lagrangian coordinate''. One has
\begin{equation}
y=a(t)x,
\end{equation}
where $a(t)>0$ is the so-called scale factor with $\dot{a}>0$ for a growing
domain. Note, however, that our formalism also accounts for the case of a
shrinking domain $\dot{a}<0$.

Analogously to section \ref{subsec:statdom} we consider the case of a
separable random walk with a (time independent) jump length PDF $\lambda(y)
\sim\sigma^{-\mu}|y|^{-1-\mu}$ for $|y|\gg \sigma$ and a waiting time PDF
$\varphi(t)\sim(t/\tau)^{-1-\alpha}$ for $t\gg\tau$. A description of the
random walk in terms of the Lagrangian coordinate is especially well-suited,
since the kinetics can then be mapped onto the original domain, at the
expense of having to deal with a time-dependent jump length PDF. Indeed,
from probability conservation, the jump length PDF on the fixed domain
(probability per unit length to take a jump of length $x$) then reads
$\lambda(x|t)=a(t)\lambda(y=a(t)x)$.

Assuming that the Fourier transform $\lambda(k)\equiv\mathscr{F}[\lambda(y)]$
takes the asymptotic form \eqref{laksmall} for small wave numbers $k$, one
finds that the Fourier transform $\lambda(k_x|t)\equiv\mathscr{F}[\lambda(x|t)]$
with respect to the Lagrangian coordinate $x$ is given by
\begin{equation}
\label{lakxsmall}
\lambda(k_x|t)= 1-\sigma^\mu\,|k_x|^\mu/a^\mu(t)+\ldots.
\end{equation}
Using both equations \eqref{lakxsmall} and the asymptotic form \eqref{varphiusmall}
of the Laplace-transformed waiting time PDF, a formalism similar to the one
introduced in section \ref{subsec:statdom} leads to the corresponding description
in terms of an FDE for the evolution of the particle's PDF $W_0(x,t)$ on the
original domain (for details we refer to \cite{levot}),
\begin{equation}
\label{W0expEq}
\frac{\partial}{\partial t}W_0(x,t)=\frac{K_\alpha^\mu}{a^\mu(t)}\nabla^\mu_x
\,_0D_t^{1-\alpha}W_0(x,t).
\end{equation}
Note that the main difference with respect to the case of a static domain given
by result \eqref{W0Eq} is that the anomalous diffusion coefficient is now
multiplied by the time dependent prefactor $a^{-\mu}(t)$. In other words, one
has an effective, time dependent anomalous diffusion coefficient $K_{\alpha,\mathrm{eff}}^
\mu\equiv a^{-\mu}(t)K_\alpha^\mu$. In the case of a growing (shrinking) domain,
one can therefore interpret that the diffusive steps measured in Lagrangian
coordinates become shorter (larger). In general, this time-dependent diffusion
coefficient complicates the solution of equation \eqref{W0expEq} considerably.
When $\alpha=1$, an analytic solution can be obtained for any value $0<\mu\le2$,
which includes the parameter range $0<\mu<2$ describing L\'evy flights (see section
\ref{sec:pulses}). In contrast, a solution in closed form does not appear to
exist when $\alpha\neq1$. However, for $\mu=2$ a careful analysis of the moments
of the distribution suffices to unveil a drastic change in the behaviour of the
solution with respect to the case of a static domain. For this particular case,
general expressions for the Lagrangian moments $\langle x^n\rangle_0$ are available,
whence expressions for the Eulerian moments $\langle y^n\rangle_0$ immediately
follow via the relation
\begin{equation}
\langle y^n\rangle_0= a^n(t)\langle x^n\rangle_0.
\end{equation}

\subsection{Behaviour of the moments for $0<\alpha\le 1$ and $\mu=2$}

Our starting point is equation \eqref{W0expEq}, which in the present case becomes
\begin{equation}
\label{W0expEqMu=2}
\frac{\partial}{\partial t}W_0(x,t)=\frac{K_\alpha}{a^2(t)}\frac{\partial^2}{
\partial x^2}\,_0D_t^{1-\alpha}W_0(x,t)
\end{equation}
with $K_{\alpha}\equiv K^2_{\alpha}$. Moments of different order can be obtained
by multiplication with the spatial variable $x$ raised to the corresponding power
and by subsequent integration.

\subsubsection{Variance}

For a generic scale factor $a(t)$, one has the formula \cite{levot}
\begin{equation}
\label{2ndmom}
\langle x^2\rangle_0=\frac{2K_{\alpha}}{\Gamma(\alpha)}\int_0^td\tau\frac{
\tau^{\alpha-1}}{a^2(\tau)}.
\end{equation}
In the case of a power-law expansion $a(t)=(1+t/t_0)^{\gamma}$ (with $\gamma \ge 0$,
whereby $\gamma=0$ corresponds to the case of a non-growing domain), one has
\begin{equation}
\label{xhypergeom}
\langle x^2\rangle_0=2K_{\alpha}t^{\alpha}_2\tilde{F}_1\left(\alpha,2\gamma,
1+\alpha,-\frac{t}{t_0}\right),
\end{equation}
where $_2\tilde{F}_1(\cdot)$ denotes the regularised hypergeometric function.
From relation (15.3.7) on page~559 of reference \cite{Abra72} one directly finds
\begin{subequations}
\label{Asy2F1}
\begin{equation}
_2\tilde{F}_1(a,b,c,-z)\sim\frac{\Gamma(b-a)}{\Gamma(b)\Gamma(c-a)}z^{-a}+\frac{
\Gamma(a-b)}{\Gamma(a)\Gamma(c-b)}z^{-b},\,\,\, a\neq b,
\label{Asy2F1a}
\end{equation}
when $z\to\infty$. Similarly, using relation (15.3.13) on page 560 of the same
reference one obtains
\begin{equation}
_2\tilde{F}_1(a,a,c,-z)\sim\frac{z^{-a}\ln(z)}{\Gamma(a)\Gamma(c-a)} .
\label{Asy2F1b}
\end{equation}
\end{subequations}
Equation \eqref{xhypergeom} in combination with equations \eqref{Asy2F1} allows one
to conclude that the Lagrangian variance displays three different asymptotic
long-time regimes, depending on the specific values of $\alpha$ and $\gamma$,
namely \cite{levot}
\begin{subequations}
\begin{equation}
\label{second-mom-slow}
\langle x^2\rangle_0\sim\frac{2K_{\alpha}t_0^{2\gamma}}{\Gamma(\alpha)(\alpha-
2\gamma)}t^{\alpha-2\gamma}
\end{equation}
for $\gamma<\alpha/2$,
\begin{equation}
\langle x^2\rangle_0\sim\frac{2K_{\alpha}t_0^{\alpha}}{\Gamma(\alpha)}\ln\left(
\frac{t}{t_0}\right)
\end{equation}
for $\gamma=\alpha/2$, and
\begin{equation}
\langle x^2\rangle_0\sim\frac{2K_{\alpha}t_0^{\alpha}\Gamma(2\gamma-\alpha)}{
\Gamma(2\gamma)}
\end{equation}
\end{subequations}
for $\gamma>\alpha/2$. In this latter case, $\langle x^2\rangle_0$ tends to a
constant value as $t\to\infty$, implying that the Lagrangian propagator
``freezes'' as a result of the fast domain growth.

Correspondingly, in physical space one gets
\begin{subequations}
\begin{equation}
\langle y^2\rangle_0\sim\frac{2K_{\alpha}}{\Gamma(\alpha)(\alpha-2\gamma)}
t^{\alpha}
\end{equation}
for $\gamma<\alpha/2$,
\begin{equation}
\langle y^2\rangle_0\sim\frac{2K_{\alpha}}{\Gamma(\alpha)}t^{\alpha}\ln\left(
\frac{t}{t_0}\right)
\end{equation}
for $\gamma=\alpha/2$, and
\begin{equation}
\langle y^2\rangle_0\sim\frac{2K_{\alpha}t_0^{\alpha-2\gamma}\Gamma(2\gamma
-\alpha)}{\Gamma(2\gamma)}t^{2\gamma}
\end{equation}
\end{subequations}
for $\gamma>\alpha/2$.

In the first case $\gamma<\alpha/2$, the domain growth is slow enough to ensure
that the long-time behaviour of the variance is the same as in case of a static
domain, except for the fact that one has a modified effective diffusion coefficient
$K_{\alpha}\alpha/(\alpha-2\gamma)$. In contrast, for $\gamma>\alpha/2$, the
particle drift associated with the deterministic domain growth (the so-called
``Hubble drift'' in the language of cosmology) becomes so large that the
intrinsic diffusive motion of the particle only represents a negligible
perturbation, and thus $\langle y^2\rangle_0\propto t^{2\gamma}$ at sufficiently
long times. Finally, in the marginal case $\gamma=\alpha/2$ the asymptotic
variance displays the same qualitative time dependence as for a non-growing
domain, but a logarithmic correction appears.

Since a sufficiently fast power-law growth $\gamma>\alpha/2$ implies that the
long-time behaviour is essentially dominated by the Hubble drift, this will
obviously also hold true for faster types of domain growth. An interesting
example is the case of an exponential growth $a(t)=\exp(Ht)$ with $H>0$. This
yields
\begin{equation}
\langle x^2\rangle_0=2K_{\alpha}(2 H)^{-\alpha}\left[1-\frac{\Gamma(\alpha,
2Ht)}{\Gamma(\alpha)}\right]\sim2K_{\alpha}(2H)^{-\alpha},
\end{equation}
implying $\langle y^2(t)\rangle_0\propto\exp(2Ht)$, as expected.

\subsubsection{Fourth-order moment}

In contrast to the case of a static domain, on a growing domain the fourth-order
moment is related to the variance in a more intricate fashion \cite{levot},
\begin{equation}
\langle x^4\rangle_0=12K_{\alpha}\int_0^t\frac{d\tau}{a^2(\tau)}\,_0D_{\tau}^{
1-\alpha}\langle x^2(\tau)\rangle_0.
\label{x^4_0VSx^2_0}
\end{equation}
Using formula \eqref{2ndmom} and performing the corresponding fractional derivative,
one obtains
\begin{equation}
\label{4thmom}
\langle x^4\rangle_0=24\frac{(K_{\alpha})^2}{\Gamma(\alpha)}\int_0^t\frac{d\tau}{
a^2(\tau)}\,_0D_{\tau}^{-\alpha}\frac{\tau^{\alpha-1}}{a^2(\tau)}.
\end{equation}
For the specific case of a power-law growth, one has
\begin{eqnarray}
\nonumber
\langle x^4\rangle_0&=&48(K_{\alpha})^2\alpha\int_0^td\tau\tau^{2\alpha}\left(1+
\frac{\tau}{t_0}\right)^{-2\gamma}\\
\nonumber
&&\times\left[\frac{1}{\tau}\,_2\tilde{F}_1\left(\alpha,2\gamma,1+2\alpha,-\frac{
\tau}{t_0}\right)\right.\\
&&-\left.\frac{\gamma}{t_0}\,_2\tilde{F}_1\left(1+\alpha,1+2\gamma,2+2\alpha,-\frac{
\tau}{t_0}\right)\right].
\end{eqnarray}
No elementary analytic expression for the above integral appears to exist. However,
it can be evaluated numerically. Conversely, equations \eqref{Asy2F1a} and
\eqref{Asy2F1b} allow one to infer the long-time behaviour of the fourth-order
moment. As in the case of the second-order moment, three different regimes can be
distinguished. For a slow expansion $\gamma<\alpha/2$, one has
\begin{subequations}
\begin{equation}
\label{fourth-mom-slow}
\langle x^4\rangle_0\sim24(K_{\alpha})^2t_0^{4\gamma}\frac{(\alpha-\gamma)\Gamma
(\alpha-2\gamma)}{(\alpha-2\gamma)\Gamma(\alpha)\Gamma(1+2\alpha-2\gamma)}t^{2
\alpha-4\gamma}.
\end{equation}
In the marginal case $\gamma=\alpha/2$, we find
\begin{equation}
\langle x^4\rangle_0\sim\frac{12(K_{\alpha})^2t_0^{2\alpha}}{[\Gamma(\alpha)]^2}
\left[\log\left(\frac{t}{t_0}\right)\right]^2.
\end{equation}
\end{subequations}
Finally, when $\gamma>\alpha/2$, since $\langle x^4\rangle_0\propto\int_0^{t/t_0}dz
z^{\alpha-2\gamma-1}$, in the long-time limit $\langle x^4\rangle_0$ tends to a
constant value (to be determined numerically). As was the case for the variance,
the Lagrangian fourth-order moment tends to a constant value. In fact, it can be
proven that whenever a ``freeze-out'' of the variance takes place, it propagates to
all even-order moments (to this end, one can e.g. take  equations (62) and (63) in reference \cite{levot}
as a starting point). Of course, this can only mean that the Lagrangian propagator tends
to a stationary profile as $t\to\infty$.

For the sake of completeness, we also give here the result for an exponential
growth $a(t)=e^{Ht}$ with $H>0$. One finds
\begin{equation}
\langle x^4\rangle_0=\frac{24(K_{\alpha})^2}{\Gamma(2\alpha)}\int_0^td\tau\exp(
-3H \tau)\tau^{2\alpha-1}\,_0F_1\left(\frac{1}{2}+\alpha,\frac{H^2\tau^2}{4}\right),
\end{equation}
where $_0F_1(\cdot)$ stands for the confluent hypergeometric function. The integral
on the right hand side remains well-defined in the limit $t\to\infty$ and the
asymptotic value of the fourth-order moment is
\begin{equation}
\label{4th-mom-exp}
\langle x^4\rangle_0(t\to\infty)\sim3\times8^{1-\alpha}(K_{\alpha})^2 H^{-2\alpha}.
\end{equation}

\subsubsection{Kurtosis}

Let us briefly discuss the behaviour of the kurtosis \eqref{kurtosis} on
the basis of the results for the fourth-order moment. To start with, note
that this quantity refers to the form of the propagator, and is therefore
scale invariant as long as the domain growth is uniform (the only case we
consider throughout the present work). Thus,
\begin{equation}
\label{kurtosis2}
\beta_2=\frac{\langle(y-\langle y\rangle)^4\rangle}{\langle(y-\langle y\rangle)^2
\rangle^2}=\frac{\langle(x-\langle x\rangle)^4\rangle}{\langle(x-\langle x\rangle)^2
\rangle^2},
\end{equation}
i.e., one may indistinctively use Eulerian or Lagrangian coordinates for the
computation of the kurtosis. For a symmetric walk, this gives
\begin{equation}
\label{sym-kurto}
\beta_2=\frac{\langle y^4\rangle_0}{\langle y^2\rangle_0^2}=\frac{\langle x^4
\rangle_0}{\langle x^2\rangle_0^2} .
\end{equation}
The main novelty introduced by the domain growth is a time-dependent kurtosis
in the subdiffusive case $0<\alpha<1$ (in the normal diffusive case $\alpha=1$,
the kurtosis remains a stationary quantity). This implies that the propagator
for the present case cannot be obtained by a simple rescaling of the propagator
referring to a static domain. More precisely, the difference $\beta_2^{\mathrm{
static}}-\beta_2$ grows in time and attains a maximum value $\beta_2^{\mathrm{
static}}-\beta_2^\infty$ as $t\to\infty$.

For the special case of the power-law growth studied previously, the behaviour
of the hypergeometric functions in the case $\gamma\le\alpha/2$ yields
(cf.~equations \eqref{second-mom-slow} and \eqref{fourth-mom-slow})
\begin{equation}
\beta_2^\infty=\frac{3\Gamma(\alpha)\Gamma(1+\alpha-2\gamma)}{\Gamma(2\alpha
-2\gamma)}.
\label{Beta2Inf_PL}
\end{equation}
In particular, this implies $\beta_2^\infty>3$ for $\gamma<\alpha/2$ and $\beta_2
^\infty=3$ for $\gamma=\alpha/2$. Starting from the case $\gamma=0$ of a
non-growing domain and increasing $\gamma$, as one approaches $\gamma=\alpha/2$
from below the final value of kurtosis decreases monotonically, and the form of
the final propagator increasingly resembles a Gaussian. In fact, when $\gamma=
\alpha/2$ the stationary kurtosis takes the value 3, i.e., that of a Gaussian
PDF. In this sense, even though the solution of equation \eqref{W0expEqMu=2} remains non-Gaussian in this case, one may still speak of Gaussian-like behaviour.

The physical origin of the time decrease of the kurtosis observed in the case $\alpha<1$
with $0<\gamma\le \alpha/2$ (and also for $\gamma>\alpha/2$, see below) is intriguing;  for $\alpha=1$, equation \eqref{W0expEqMu=2} describes scaled Brownian motion, and the propagator remains strictly Gaussian at all times. In contrast, for $\alpha<1$ the propagator is fat-tailed at short times; however, as time goes by,
the effect of the Hubble drift on these fat tails of the distribution appears to become stronger than in the
central part. This could explain that for $\alpha<1$ the kurtosis takes values which are increasingly close to the Gaussian value $\beta_2=3$ in the course of time.  We have no physical explanation for the Gaussian-like behaviour observed when $\gamma = \alpha/2 $, other than the fact that this particular value separates the diffusion-dominated regime from the regime dominated by the domain growth \cite{levot}.

\begin{figure}
\centering
\includegraphics[width=8cm]{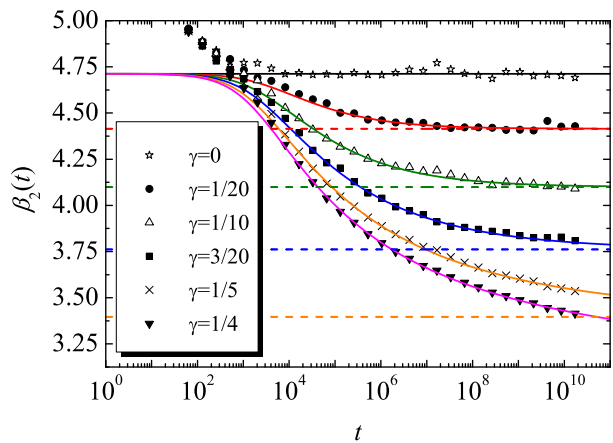}
\caption{Kurtosis for a subdiffusive random walk with $\alpha=1/2$
and $K_{\alpha}=1/2$ on an evolving domain with power-law scale factor $a(t)=(
1+t/t_0)^{\gamma}$, for $t_0=10^3$ and $\gamma=0$, $1/20$, $1/10$, $3/20$, $1/5$,
and $1/4$. Symbols represent simulations results with $10^6$ realisations.
Numerical solutions are represented by solid lines. The value of $\beta_2(t)$
was computed by numerical integration of the Lagrangian fourth-order moment
and subsequent division by the analytical expression of the squared Lagrangian
second-order moment. Horizontal dashed lines represent the value of $\beta_2^{
\infty}$ for each value of $\gamma$ according to equation \eqref{Beta2Inf_PL}.
The asymptotic value for $\gamma=1/4$, $\beta_2^{\infty}=3$, is not shown.}
\label{Fig:Beta2_Slow}
\end{figure}

In figure~\ref{Fig:Beta2_Slow} we show simulations results for a power-law
expansion $a(t)=(1+t/t_0)^{\gamma}$ with $t_0=10^3$ and different values
of $\gamma \leq\alpha/2$. The curves for $\beta_2(t)$ can be seen to
approach the value of $\beta_2^{\infty}$ given by result \eqref{Beta2Inf_PL}. In
particular, the decrease of $\beta_2^{\infty}$ with increasing $\gamma$
predicted by equation \eqref{Beta2Inf_PL} is confirmed. The theoretical results
compare very favourably with Monte-Carlo simulations at times sufficiently long
to reach the diffusive regime.

\begin{figure}
\centering
{\includegraphics[width=7.6cm]{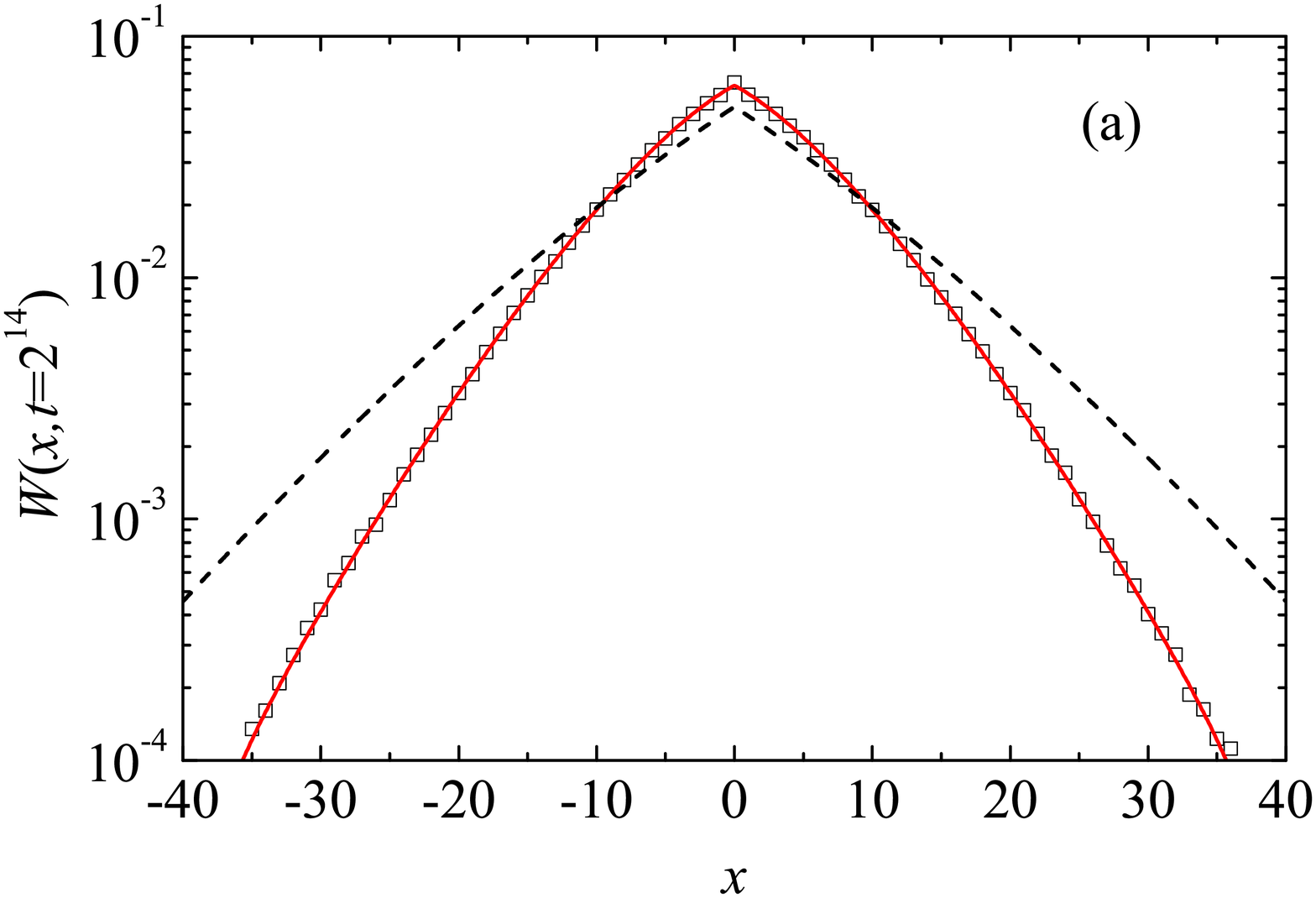}}
{\includegraphics[width=7.6cm]{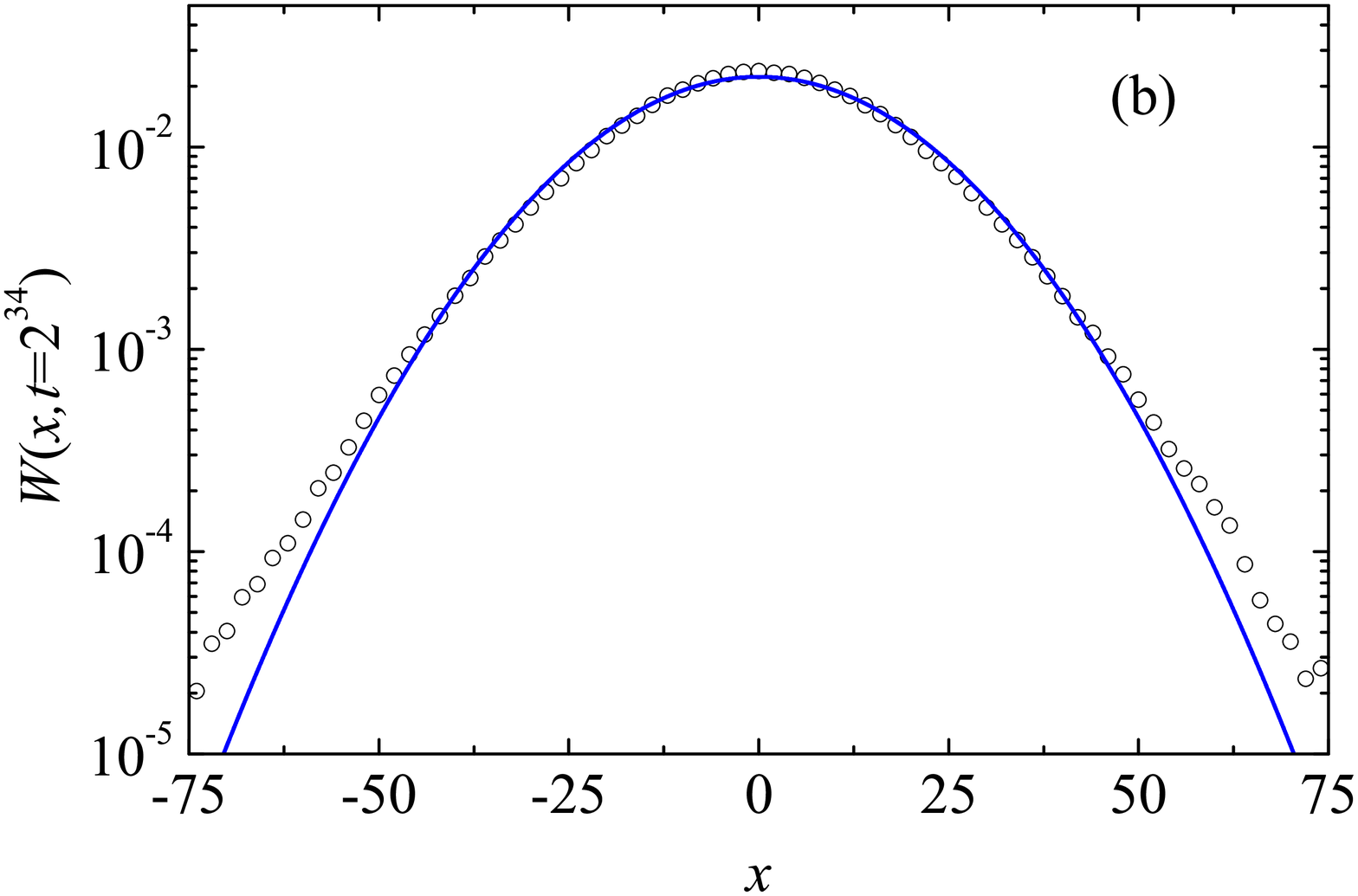}}
\caption{Logarithmic representation of the Lagrangian propagator $W(x,t)$ for a
subdiffusive particle with $\alpha=1/2$ and $K_{\alpha}=1/2$. The domain growth
is given by the power-law scale factor $a(t)=(1+t/10^3)^{1/4}$. In panel (a) we
show the simulations results (empty squares) for the propagator at time $t=2^{14}$,
together with numerical solution of equation \eqref{W0expEqMu=2} obtained via a fractional
finite-difference method \cite{yuste_weighted} with spatial discretisation $\Delta x =0.1$ and time
discretisation $\Delta t=0.1$ (solid line). The dashed line represents the exact
solution for the static case at $t=2^{14}$. In panel (b) we show the propagator
at $t=2^{34}$ obtained from simulations (empty circles), and we compare it with
a Gaussian whose variance is taken to be that of the real distribution.}
\label{Fig:W_g025}
\end{figure}

In figure~\ref{Fig:W_g025} we illustrate how the form of the propagator
changes as its kurtosis evolves in time. We show the Lagrangian propagator for a
power-law expansion $a(t)=(1+t/t_0)^{\gamma}$ with $t_0=10^3$ and $\gamma=\alpha
/2=1/4$ at two different times. In panel (a), we display the propagator at a
comparatively short time ($t=2^{14}$). We have been able to solve the FDE computationally
up to this time (details of the algorithm can be found in reference \cite{yuste_weighted}). The propagator
is still quite pointy at $x=0$ and thus not too different from the shape on a static domain ($\gamma=0$). In more
quantitative terms, at $t=2^{14}$ the kurtosis of $W(x,t)$ is $\beta_2\left(2^{14}
\right)\simeq 4.22$, whereas on a static domain one would have $\beta_2^{\mathrm{
static}}\approx4.71$. However, at the later time $t=2^{34}$---see panel (b)---our
numerical Monte-Carlo simulations reveal that the sharp peak at $x=0$ evolves into
a bell-shaped hump. This signals the evolution of the propagator towards the
Gaussian-like state reached as $t\to\infty$. However, the kurtosis $\beta_2\left(
t=2^{34}\right)\simeq3.42$ at this time is still clearly distinguishable from the
final value $\beta_2^{\infty}=3$.

\begin{figure}
\centering
\includegraphics[width=8cm]{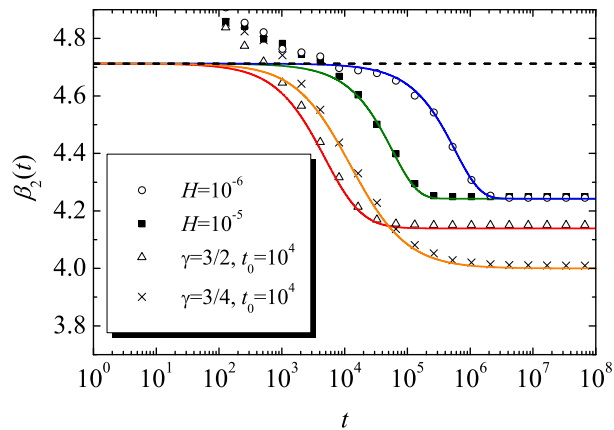}
\caption{Kurtosis for a subdiffusive random walk with $\alpha=1/2$ and $K_{\alpha}
=1/2$ on fast growing domains ($\gamma>\alpha/2$ for power-law expansions  and $H>0$ for exponential expansions). The two upper curves correspond to an exponential
scale factor $a(t)=\exp(Ht)$, whose logarithmic derivative is, respectively, $H
=10^{-5}$ and $10^{-6}$. The two bottom curves correspond to a power-law scale
factor $a(t)=(1+t/t_0)^{\gamma}$ with $t_0=10^4$ and $\gamma=3/2$ and $3/4$, from
top to bottom at $t=10^8$. Symbols represent simulation results from $10^6$
realisations. Numerical solutions are depicted by solid lines. The value of
$\beta_2(t)$ was computed by numerical integration of the Lagrangian fourth-order
moment and subsequent division by the analytical expression of the squared
Lagrangian second-order moment. The dashed line depicts the value of $\beta_2^{
\mathrm{static}}\simeq4.71$ (cf.~equation \eqref{Beta2_Static}).}
\label{Fig:Beta2_Fast}
\end{figure}

In contrast to the case $\gamma\leq\alpha/2$ for a sufficiently fast domain
growth ($\gamma>\alpha/2$) the final value of the kurtosis $\beta_2^\infty$
increases as $\gamma$ grows. As shown in reference \cite{levot}, in this
$\gamma$-regime the Lagrangian propagator eventually freezes and its final
form is always leptokurtic ($\beta_2^\infty>3$). This also implies the
long-time freeze-out of all the associated moments, which tend to finite
values as $t\to\infty$. In particular, this holds for the second- and the
fourth-order moments, from which the kurtosis is computed. As explained in
reference \cite{levot}, the physical reason for the observed freeze-out is the irrelevance
of the diffusive spreading with respect to the Hubble drift at
sufficiently long times. As a result of this, after a characteristic time
$t_{\mathrm{char}}>t_0$ (see figure \ref{Fig:Beta2_Fast}), changes in the form of the
Lagrangian propagator and in the associated kurtosis become negligibly small, and
the monotonic time decrease of the kurtosis saturates at a value $\beta_2^{\infty}\in(3,\beta_2^{
\mathrm{static}})$. Of course, $t_{\mathrm{char}}$ will depend on both $t_0$
and $\gamma$. For larger values of $\gamma$, one expects a decrease of $t_{
\mathrm{char}}$, since the Hubble drift becomes dominant with respect to
the diffusive spreading at earlier times, and hence the saturation in
Lagrangian space becomes faster. This entails a stronger memory of the
tailedness displayed by the early-time propagator, and therefore a larger
$\gamma$ leads to a larger $\beta_2^{\infty}$. Note that this is just the
opposite of what happens when $0<\gamma\leq\alpha/2$. Our findings for the
case $\gamma>\alpha/2$ are fully confirmed by the results shown in figure
\ref{Fig:Beta2_Fast}, which displays a comparison between theory and
simulations in the long-time regime.

Summarising, Gaussian-like behaviour is only observed when $\gamma=\alpha/2$.
For any other value the kurtosis of the final distribution is always $>3$, i.e.,
the final PDF is leptokurtic and a strong signature of the early-time PDF persists
for arbitrarily long times. Note, however, that the asymptotic value $\beta_2^\infty$
does not depend on the characteristic time $t_0$, which only has an influence on the
transient behaviour.

Of course, in the exponential case $a(t)=e^{Ht}$ with $H>0$, a freeze-out of
the Lagrangian propagator also takes place. It is, however, striking that $\beta_2^{
\infty}=3\times2^{1-\alpha}$, regardless of the value of $H$. For any $\alpha<1$ one
has $\beta_2^\infty>\beta_2^{\mathrm{Gaussian}}\equiv3$, but $\beta_2^\infty<\beta_2^{
\mathrm{static}}\equiv3\Gamma(\alpha)\Gamma(1+\alpha)/\Gamma(2\alpha)$---see
equation \eqref{Beta2_Static}.

Results for two different values of $H>0$ are displayed in figure
\ref{Fig:Beta2_Fast}. The kurtosis $\beta_2(t)$ in this case is seen to approach
$\beta_2^{\infty}$ at a time of the order of $1/H$. As already anticipated
$\beta_2^{\infty}$ does not depend on $H$.

Of interest is also the behaviour in the case of exponential contraction,
i.e., $a(t)=e^{Ht}$ with $H<0$. In this case it is easier to work in physical
coordinates. In the limit $t\to\infty$ one has the asymptotic behaviour
\begin{subequations}
\begin{eqnarray}
\label{second-phys-mom-exp}
\langle y^2\rangle_0 \sim \frac{ K_\alpha t^{\alpha - 1} }{ |H| \Gamma(\alpha)}, &&\\
\label{fourth-phys-mom-exp}
\langle y^4\rangle_0\sim\frac{3\times2^{1-\alpha}(K_{\alpha})^2}{|H|^{1+\alpha}
\Gamma(\alpha)}t^{\alpha-1}. &&
\end{eqnarray}
\end{subequations}
whence
\begin{equation}
\beta_2\sim3\times2^{1-\alpha}|H|^{1-\alpha}t^{1-\alpha}.
\end{equation}
For further details of the calculations of the moments, including the use of
Tauberian theorems, we refer to reference \cite{levot}.

As expected, when $\alpha=1$ the kurtosis is equal to three, whereas for
$\alpha<1$ the kurtosis always grows in time (in this case, it is proportional
to $t^{1-\alpha}$).

Other types of contraction can also be considered. The case of power-law
contraction, i.e., $a(t)=(1+t/t_0)^\gamma$ with $\gamma<0$, is also covered by
our formalism. The kurtosis $\beta_2(t)$ increases in time, until a limiting
value $\beta_2^\infty$ is attained. This limiting value is still given by
equation \eqref{Beta2Inf_PL}, which also holds for $\gamma<0$. For a given
$\alpha<1$, the value of $\beta_2^\infty$ grows with increasing
$|\gamma|$.

We close this subsection by noting that the observed time increase of $\beta_2$ for shrinking domains
is likely to be related to the inversion of the direction of the Hubble drift with respect to
the case of a growing domain (in uniformly shrinking domains, the Hubble drift tends to bring any two physical points closer to each other, whereas it tends to separate them in a uniformly growing domain). We actually
conjecture that beyond the two cases with exponential and power-law scale factor studied here, the kurtosis of the propagator generated by subdiffusive walks with $\alpha<1$ displays a time decrease (increase) on any uniformly growing (shrinking) infinite domain.

\subsection{L\'evy flights}
\label{subsec:Levy}

As already anticipated, in the case of L\'evy flights ($\alpha=1$ and $\mu<2$)
the propagator $W_0(x,t)$ associated with equation \eqref{W0expEq} can be
explicitly obtained. One has
\begin{equation}
\label{Levyprop}
W_0(x,t)=\mathsf{L}_{\mu}\left(x;K_1^{\mu}\int_0^ta^{-\mu}(u)du\right),
\end{equation}
where
\begin{equation}
\mathsf{L}_{\mu}(x;\sigma_L^{\mu})=\mathcal{F}^{-1}\left[\exp\left(-|k\sigma_L|
^{\mu}\right)\right]
\end{equation}
is a symmetric L\'evy-stable density with exponent $\mu$ and scale factor
$\sigma_L$. Note that for $\mu=2$, this L\'evy density becomes a Gaussian
PDF with standard deviation $\sigma=\sqrt{2}\sigma_L$.

By definition, the second-order moment of a L\'evy flight is infinite. However,
one can define a typical width as $w_\mu(t)\equiv C_{\mu}\sigma_L(t)$, where
$C_{\mu}$ is a constant chosen in such a way that
\begin{equation}
\label{tip-width-Levy}
\int_0^{w_{\mu}(t)}W_0(x,t)dx=P/2
\end{equation}
holds, where $P$ is a predetermined probability.

The typical width $w_\mu$ will be seen to play an important role when addressing
the kinetics of mixing of two initially localised L\'evy pulses evolving on a
growing domain (cf.~section \ref{sec:pulses}).

\subsection{Fractional diffusion equation in physical coordinates}

It is possible to obtain the PDF $W_0^*(y,t)$ to find a particle inside the
interval $[y,y+dy]$ at time $t$ by noting that it is related to the PDF
$W(x,t)$ for finding the particle in the corresponding interval $[x,x+dx]$,
where $x=y/a(t)$. One has \cite{levot}
\begin{equation}
\label{relWyWx}
W_0^*(y,t)=\frac{W_0(x=y/a,t)}{a(t)}.
\end{equation}
This implies the three relations
\begin{equation}
\label{rel1}
\left.\frac{\partial W_0}{\partial t}\right|_{x}=a \left.\frac{\partial W_0^*}{
\partial t}\right|_{y}+\dot{a}\left.\frac{\partial(y W_0^*)}{\partial y}\right|_t,
\end{equation}
\begin{equation}
\label{rel2}
\partial W_0/\partial x|_t=a^2\partial W_0^*/\partial y|_t,
\end{equation}
and
\begin{equation}
\label{rel3}
\nabla^\mu_xW_0(x,t)=a^{1+\mu}\nabla^\mu_yW_0^*(y,t).
\end{equation}
Inserting relations \eqref{rel1} to \eqref{rel3} into equation \eqref{W0expEq} one
eventually finds the sought equation
\begin{equation}
\fl
\label{EcnodriftWasty}
\frac{\partial W_0^*(y,t)}{\partial t}=-\frac{\dot{a}(t)}{a(t)}\frac{\partial}{
\partial y }\left[yW_0^*(y,t)\right]+K_\alpha^{\mu}\nabla^\mu_y a(t)^{-1}\left\{
\,_0D_t^{1-\alpha}\left[a(t)W_0^*(a(t)x,t)\right]\right\}_{x\to y/a(t)}.
\end{equation}
In the case $\mu=2$ it is possible to directly obtain the behaviour of the
physical moments $\langle y^n(t)\rangle\equiv\int_{-\infty}^\infty y^nW_0^*(
y,t)dy$ by multiplication of equation \eqref{EcnodriftWasty} with $y^n$ and
subsequent integration over $y$. This yields a hierarchy of differential
equations for the physical moments.

\section{CTRW in a velocity field}
\label{sec:biased}

The diffusion of a particle in a uniform velocity field can be regarded as the
motion of a random walker dragged by a fluid flowing with velocity $\vec v$
with respect the laboratory reference frame $\mathcal{S}_L$ (for simplicity,
the velocity $\vec v$ will hereafter be assumed to be stationary unless
otherwise specified). An example would be a frog performing random jumps with
statistically distributed waiting times on a wooden log that is longitudinally
floating downstream on a river. On a more microscopic scale one could imagine
a tracer particle subdiffusing in a hydrogel that itself is slowly streamed in a fluidic
device.

Let us now introduce a second reference frame $\mathcal{S}_0$ in which the
deterministic contribution of the velocity field is subtracted from the overall
particle motion, i.e., a frame which follows the fluid that drags the particle
along. Clearly, $\mathcal{S}_L$ moves with velocity $-\vec v$ with respect to
$\mathcal{S}_0$. Let us respectively denote by $W_0(\vec y,t)$ and $W(\vec y,t)$
the walker's PDF in $\mathcal{S}_0$ and $\mathcal{S}_L$. On a static domain the
relation between both PDFs will be given by the Galilean transformation $W(\vec
y,t)=W_0(\vec{y}-\vec{v}t,t)$. In particular one has
\begin{equation}
\label{bias-no-bias}
W(y,t)=W_0(y-vt,t)
\end{equation}
in one dimension. In Fourier-Laplace space equation \eqref{bias-no-bias} becomes
\begin{equation}
\label{WGaku}
W(k,u)=W_0(k,u+ikv).
\end{equation}

In the case of a growing domain the particle is not only advected by the
velocity field but also experiences an additional drift as it is dragged by
the physical medium (i.e., the aforementioned Hubble drift). Recalling the
previous example of a hopping frog on a log floating downstream, one might
wonder what the equation of motion of the frog would be if the log were
replaced by a linear rubber strip that expands uniformly with a certain scale
factor. Before providing the answer to this question, we will first address
the simpler case of a static domain, as done in section \ref{sec:unbiased}.

\subsection{Static domain}

From equations \eqref{WGaku} and \eqref{W0kuljr} one finds
\begin{equation}
\label{FLWku}
W(k,u)=\frac{W(k,0)}{u+ivk+K_\alpha^\mu|k|^\mu(u+ivk)^{1-\alpha}},
\end{equation}
or, equivalently,
\begin{equation}
(u+ivk)W(k,u)-W(k,0)=-K_\alpha^\mu|k|^\mu(u+ivk)^{1-\alpha} W(k,u),
\end{equation}
whence the Galilei-invariant (GI) advection-diffusion equation
\begin{equation}
\label{EcuWxtljr}
\left(\frac{\partial}{\partial t}+v\frac{\partial}{\partial y}\right)W(y,t)=
K_\alpha^\mu\nabla^\mu_y\left(\frac{\partial}{\partial t}+v\frac{\partial}{
\partial y}\right)^{1-\alpha}W(y,t)
\end{equation}
follows. The operator $\left(\partial_t+v\,\partial_ y\right)^{1-\alpha}$
is the fractional material derivative introduced by Sokolov and Metzler
\cite{Sokolov2003}. It is defined by
\begin{equation}
\mathscr{F}\mathscr{L}\left[\left(\frac{\partial}{\partial t}+v\frac{\partial}{
\partial y}\right)^{1-\alpha}W(y,t)\right]=(u+ivk)^{1-\alpha}W(k,u).
\end{equation}
From this expression and from relation \eqref{GLLaplace} one can see that the
fractional material derivative reduces to the GL derivative if $v=0$. As mentioned
in the Introduction, the GI fractional advection-diffusion equation \eqref{EcuWxtljr}
for $\mu=2$ was recently obtained by Cairoli {\it et al.} \cite{Cairoli2018}. While the
standard material derivative, corresponding to the limit $\alpha=1$ reflects
the GI of a standard physical system, the power $(1-\alpha)$ reflects the
spatio-temporal coupling in a waiting time-random walk scenario with a
constant relocation speed $v$.

The fractional derivative in direct (position-time) space is \cite{Uchaikin2014}
\begin{equation}
\label{MatDer}
\left(\frac{\partial}{\partial t}+v\frac{\partial}{\partial y}\right)^{1-\alpha}
=\frac{1}{\Gamma(\alpha)}\left(\frac{\partial}{\partial t}+v\frac{\partial}{\partial
y} \right)\int_0^t d\tau\frac{ W(y-v(t-\tau),\tau)}{(t-\tau)^{1-\alpha}}.
\end{equation}
Here, it is assumed that $W(k,t)$ satisfies the condition \eqref{GLeqRLcond} of
good behaviour at $t=0$.\footnote{Equation \eqref{MatDer} is slightly different
from that in \cite{Uchaikin2014}, since in that reference the Fourier transform
is defined as in equation \eqref{FTdefi} but with the replacement $k\to -k$).}

The fractional material derivative can be rewritten in terms of the RL integral
in the convenient form
\begin{eqnarray}
\label{MatDerbis}
\left(\frac{\partial}{\partial t}+v\frac{\partial}{\partial y}\right)^{1-\alpha}
W(y,t)&=&\left(\frac{\partial}{\partial t}+v\frac{\partial}{\partial y}\right)
\left[\,_0D_t^{-\alpha} W(y+vt,t)\right]_{y\to y-vt}\nonumber\\
&=&\left(\frac{\partial}{\partial t}+v\frac{\partial}{\partial y}\right)\left[
\,_0D_t^{-\alpha} W_0(y,t)\right]_{y\to y-vt}.
\end{eqnarray}
In this way the GI advection-diffusion equation \eqref{EcuWxtljr} becomes
\begin{equation}
\label{EcWxtSh}
\left(\frac{\partial}{\partial t}+v\frac{\partial}{\partial y}\right)W(y,t)=
K_\alpha^\mu\nabla^\mu_y\left(\frac{\partial}{\partial t}+v\frac{\partial}{
\partial y} \right)\left[\,_0D_t^{-\alpha} W_0(y,t)\right]_{y\to y-vt}.
\end{equation}
Note that in the above expression the time fractional derivative is applied to
$W_0$, which corresponds to the particle distribution in the reference frame
$\mathcal{S}_0$. This frame moves by a distance $vt$ during the time $t$. As a
result of this the evaluation point for the derivative is shifted by the same
quantity.

Finally, it should be noted that the propagator \eqref{FLWku} and the corresponding FDAE \eqref{EcWxtSh} differ from those considered in references \cite{Metzler1998} and \cite{Fa2011}. In particular, for $\mu = 2 $, the propagator studied in these works can reach non-physical negative values with moments that are only correct up to order two.

\subsection{Growing domain}

We continue to assume that the particle is subject to the influence of a constant
velocity field (the velocity measured with respect to the laboratory frame $\mathcal{S}_L$
is $v$). As in the case of a symmetric walk it is convenient to work in Lagrangian
coordinates. If the domain expands uniformly with the scale factor $a(t)$, the
Lagrangian distance travelled by the reference frame $\mathcal{S}_0$ (the frame
moving with velocity $v$ with respect to $\mathcal{S}_L$) during the time $t$ is
\begin{equation}
\Lambda(t)=\int_0^t\frac{v}{a(t')}\,dt'\equiv \int_0^t\nu(t')\,dt'\equiv vT(t).
\end{equation}
Here, the transformed time $T(t)$ replaces the physical time $t$ as a result of
the difference between the Lagrangian distance $\Lambda(t)$ travelled by $\mathcal{S}_0$
and the distance $vt$ that it would cover on a
static domain.\footnote{Notice that in cosmology, $\Lambda$,  $T$ and $\nu$  are known
as ``comoving distance'', ``conformal time'', and ``comoving velocity'',  respectively \cite{Ryden2003}.}
For instance, an exponential growth $a(t)=\exp(Ht)$ with $H>0$ yields,
\begin{equation}
\label{trans-time-exp}
T(t)=\frac{1-\exp(-Ht)}{H},
\end{equation}
whereas a power-law growth $a(t)=(1+t/t_0)^\gamma$ gives
\begin{subequations}
\label{trans-time-pot}
\begin{equation}
\label{trans-time-pot-1}
T(t)=\frac{t_0}{\gamma-1}\left[1-\left(\frac{t}{t_0}\right)^{1-\gamma}\right],
\qquad\gamma\neq1
\end{equation}
and
\begin{equation}
\label{trans-time-pot-2}
T(t)=t_0 \ln\left(1+\frac{t}{t_0}\right),\qquad\gamma=1.
\end{equation}
\end{subequations}
Thus, for a sufficiently fast growth (exponential or power-law with $\gamma>1$)
one has an asymptotic finite value $T_\infty\equiv T(t\to\infty)$, and
correspondingly the asymptotic Lagrangian distance $\Lambda_\infty\equiv\Lambda(
t\to\infty)$ is also \emph{finite}.

\begin{figure}
\centering
\includegraphics[width=8cm]{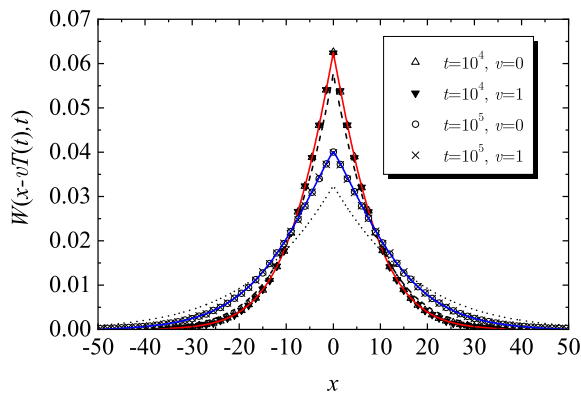}
\caption{Lagrangian propagators for a subdiffusive random walk with $\mu=2$,
$\alpha=1/2$, and $K_{\alpha}=1/2$. The domain growth is given by a power-law
scale factor $a(t)=(1+t/10^3)^{1/8}$. We display cases with $v=0$ (zero field)
and $v=1$. Symbols depict simulations results at times $t=10^4$ and $t=10^5$
for the case $v=0$ and $v=1$ (see legend). Solid lines represent numerical
solutions of equation \eqref{W0expEqMu=2} at these times (the temporal and
spatial steps are, respectively, $\Delta t=t/10^5$ and $\Delta x =0.1$). The
broken lines are the exact solutions at $t=10^4$ (dashed) and $t=10^5$ (dotted)
for a static domain.}
\label{Fig:W_Velocity}
\end{figure}

The PDF $W(x,t)$ in the laboratory frame $\mathcal{S}_L$ is just the PDF $W_0(z,t)$ in
the comoving reference frame $\mathcal{S}_0$, but with the shifted position $z=
x-\Lambda(t)$. In other words,
\begin{equation}
\label{WGaExp}
W(x,t)=W_0(x-\Lambda(t),t).
\end{equation}
In figure \ref{Fig:W_Velocity} the above relation is illustrated for a
stretching domain with $a(t)=(1+t/10^3)^{1/8}$ by comparing simulations
results for $W_0$ and $W$. The Lagrangian propagators are also compared
with the corresponding ones for the case of a static domain. Note that the
width of the propagator is decreased with respect to the static case, since
the jump length measured in Lagrangian coordinates is divided by $a(t)$.

Now, as we already know from section \ref{sec:unbiased}, $W_0(x,t)$ satisfies
relation \eqref{W0expEq}. Since $\Lambda(t)$ may be a complicated function of $t$,
in general no relationship between the Fourier-Laplace transforms of $W(x,t)$ and $W_0(x,t)$
similar to equation \eqref{WGaku} can be obtained from result \eqref{WGaExp}.
Therefore,  we can no longer use the straightforward procedure of section \ref{sec:unbiased}
to derive the FDAE that we seek. For this reason, we will follow a
different path involving equations \eqref{W0expEq}, \eqref{EcWxtSh}, and
\eqref{WGaExp}.

In view of equations \eqref{EcWxtSh} and \eqref{W0expEq} a reasonable guess for
the FDAE is
\begin{equation}
\label{ecuWexp}
\left(\frac{\partial}{\partial t}+\nu\frac{\partial}{\partial x}\right)W(x,t)=
\frac{K_\alpha^\mu}{a^\mu(t)}\nabla^\mu_x\left[_0D_t^{1-\alpha}W_0(x,t)\right]_{
x\to x-\Lambda(t)},
\end{equation}
since $\Lambda(t)$ and $vt$ in equations \eqref{ecuWexp} and \eqref{EcWxtSh} are,
respectively, the displacement in the laboratory frame $\mathcal{S}_L$ of the
comoving frame $\mathcal{S}_0$ during the time $t$. We confirm that equation
\eqref{ecuWexp} is indeed the correct FDAE by showing that $W(x,t)$, as given
by equation \eqref{WGaExp}, satisfies relation \eqref{ecuWexp}. First, let
us evaluate the left-hand side of equation \eqref{ecuWexp},
\begin{eqnarray}
\left(\frac{\partial}{\partial t}+\nu\frac{\partial}{\partial x}\right)W(x,t)
=\left(\frac{\partial}{\partial t}+\nu\frac{\partial}{\partial x}\right)W_0(x
-\Lambda(t),t)=\left.\frac{\partial W_0}{\partial t}\right|_{x-\Lambda(t)}.
\label{leftEq}
\end{eqnarray}
Next, we evaluate the right-hand side of equation \eqref{ecuWexp},
\begin{eqnarray}
\nonumber
\frac{K_\alpha^\mu}{a^\mu(t)}\nabla^\mu_x\left[_0D_t^{1-\alpha}W_0(x,t)\right]_{
x\to x-\Lambda(t)}&=&\left[\frac{K_\alpha^\mu}{a^\mu(t)}\nabla^\mu_x\,_0D_t^{1-
\alpha}W_0(x,t)\right]_{x \to x-\Lambda(t)}\\
&=&\left.\frac{\partial W_0}{\partial t}\right|_{x-\Lambda(t)},
\label{rightEq}
\end{eqnarray}
where, in the last step, equation \eqref{W0expEq} was taken into account. Comparing
equation \eqref{leftEq} with \eqref{rightEq} we conclude that $W(x,t)=W_0(x-\Lambda
(t),t)$ indeed satisfies relation \eqref{ecuWexp}. Defining
\begin{equation}
\fl
\label{matderOp}
\left(\frac{\partial}{\partial t}+\nu\frac{\partial}{\partial x}\right)^{1-\alpha}
W(x,t)\equiv\left(\frac{\partial}{\partial t}+\nu\frac{\partial}{\partial x}\right)
\frac{1}{\Gamma (\alpha)}\int_0^td\tau\frac{W\left(x+\Lambda(\tau)-\Lambda(t),\tau
\right)}{(t-\tau)^{1-\alpha}},
\end{equation}
the FDAE \eqref{EcWxtSh} for a uniformly growing domain can be rewritten in a way
similar to equation \eqref{EcuWxtljr},
\begin{equation}
\label{EcuWxExp}
\left(\frac{\partial}{\partial t}+\nu\frac{\partial}{\partial x}\right)W(x,t)=
\frac{K_\alpha^\mu}{a^\mu(t)}\nabla^\mu_x\left(\frac{\partial}{\partial t}+\nu
\frac{\partial }{\partial x}\right)^{1-\alpha} W(x,t).
\end{equation}
Equation \eqref{EcuWxExp}, or equivalently equation \eqref{ecuWexp} (see just below), is one
of the main results of this paper.

Comparing equations \eqref{ecuWexp} and \eqref{EcuWxExp} one can see that both
are equivalent if
\begin{equation}
\label{equiv1}
\left(\frac{\partial}{\partial t}+\nu\frac{\partial}{\partial x}\right)^{1-\alpha}
W(x,t)=\left[_0D_t^{1-\alpha}W_0(x,t)\right]_{x\to x-\Lambda(t)}.
\end{equation}
In order to prove this we first note that equation \eqref{matderOp} can be
rewritten as (see equation \eqref{RLIntegral})
\begin{equation}
\label{equiv2}
\left(\frac{\partial}{\partial t}+\nu\frac{\partial}{\partial x}\right)^{1-\alpha}
W(x,t)=\left(\frac{\partial}{\partial t}+\nu\frac{\partial}{\partial x}\right)
\left[_0D_t^{-\alpha}W(x+\Lambda(t),t)\right]_{x\to x-\Lambda(t)}.
\end{equation}
Conversely, for any differentiable function $G(x,t)$, one has
\begin{eqnarray}
\nonumber
\left(\frac{\partial}{\partial t}+\nu\frac{\partial}{\partial x}\right)\left[
G(x,t)\right]_{x\to x-\Lambda(t)}=\left(\frac{\partial}{\partial t}+\nu\frac{
\partial}{\partial x}\right)G(x-\Lambda(t),t)\\
=\left.\frac{\partial G}{\partial t}\right|_{x-\Lambda(t)}-\frac{\partial\Lambda
(t)}{\partial t}\left.\frac{\partial G}{\partial x}\right|_{x-\Lambda(t)}+\nu\left.
\frac{\partial G}{\partial x}\right|_{x-\Lambda(t)}=\left.\frac{\partial G}{\partial
t} \right|_{x-\Lambda(t)}.
\label{Gx12}
\end{eqnarray}
Taking $G(x,t)=\,_0D_t^{-\alpha}W(x+ \Lambda(t),t)$ and assuming that $(d/dt){_0}D_t
^{-\alpha}=\,_0D_t^{1-\alpha}=\,_0\mathcal{D}_t^{1-\alpha}$ (see equation
\eqref{GLeqRL}), one finds that the right hand side of equations \eqref{equiv1} and
\eqref{equiv2} coincide, implying that relation \eqref{equiv1} indeed holds.

Even though the velocity $v$ has been assumed to be constant, it is worth noting
that equations \eqref{ecuWexp} and \eqref{EcuWxExp} still hold when the reference
frame $\mathcal{S}_0$ moves with a time-dependent velocity with respect to
$\mathcal{S}_L$. In this case, one simply replaces $\nu=v/a(t)$ with $\nu=v(t)/a(t)$.

We close this section by deriving the FDAE in terms of the physical coordinate $y$.
The derivation proceeds along the same lines as in the field-free case (cf.~section
\ref{sec:unbiased}). The relations \eqref{relWyWx} to \eqref{rel3} for this latter
case are completely analogous to the equations relating $W^*(y,t)$ and $W(x,t)$ in
the presence of the velocity field. Indeed, one has
\begin{equation}
\label{relWyWx-nd}
W^*(y,t)=\frac{W(x=y/a,t)}{a(t)},
\end{equation}
and, correspondingly,
\begin{subequations}
\begin{equation}
\label{rel1-nd}
\left.\frac{\partial W}{\partial t}\right|_x=a\left.\frac{\partial W^*}{\partial
t}\right|_y+\dot{a}\left.\frac{\partial(yW^*)}{\partial y}\right|_t,
\end{equation}
as well as
\begin{equation}
\label{rel2-nd}
\left.\frac{\partial W}{\partial x}\right|_t=a^2\left.\frac{\partial W^*}{\partial y}\right|_t,
\end{equation}
and
\begin{equation}
\label{rel3-nd}
\nabla^\mu_x W(x,t)=a^{1+\mu}\,\nabla^\mu_yW^*(y,t).
\end{equation}
\end{subequations}
Inserting equations \eqref{relWyWx-nd} to \eqref{rel3-nd} into \eqref{ecuWexp} and
taking into account equation \eqref{WGaExp} one eventually obtains the result
\begin{equation}
\label{EcWasty}
\frac{\partial W^*}{\partial t}=-\frac{\partial}{\partial y}\left[\left(\frac{\dot{
a}}{a}y+a\nu\right)W^*(y,t)\right]+K_\alpha^\mu \nabla^\mu_y{_0}P_t^{1-\alpha}W^*(y,t),
\end{equation}
where
\begin{equation}
\label{p-operator}
_0P_t^{1-\alpha} W^*(y,t)=\frac{1}{a}\left\{\,_0D_t^{1-\alpha}\left[aW^*(ax+a
\Lambda,t)\right]\right\}_{x=(y-\Lambda)/a}.
\end{equation}
Let us once more recall that equation \eqref{EcWasty} holds for well-behaved functions
$W^*(y,t)$. Strictly speaking, the RL-fractional derivative therein must be replaced
with the GL-fractional derivative $_0{\cal D}_t^{1-\alpha}$, and consequently, $_0P_t
^{1-\alpha}$ must also be replaced with the corresponding operator, defined via the
equation
\begin{equation}
\label{p-operator-2}
_0{\cal P}_t^{1-\alpha}W^*(y,t)=\frac{1}{a}\left\{\,_0{\cal D}_t^{1-\alpha}
\left[aW^*(ax+a\Lambda,t)\right]\right\}_{x=(y-\Lambda)/a}.
\end{equation}

\subsection{Moments of the propagator}

From the exact relationship \eqref{WGaExp} one can easily find the moments
$\langle x^n\rangle$ of $W$ in terms of the moments $\langle x^n\rangle_0$
of $W_0$. One has
\begin{equation}
\label{xnx0}
\langle x^n\rangle=\int(z+\Lambda)^nW_0(z,t)=\sum_{m=0}^n {n\choose m}
\Lambda^{n-m}\langle x^m\rangle_0.
\end{equation}
Restricting ourselves to the case $\mu=2$ and $0<\alpha\le1$ (subdiffusive case), the second- and
fourth-order moments of the PDF are respectively given by equations \eqref{2ndmom}
and \eqref{4thmom}. By symmetry, odd moments vanish trivially, $\langle x^{2n+1}\rangle_0\equiv
0$. Taking all this into account in equation
\eqref{xnx0} we find explicit expressions for the first four Lagrangian moments:
\begin{subequations}
\begin{eqnarray}
\label{x1WW}
\langle x\rangle =\Lambda(t), &&\\
\label{x2WW}
\langle x^2\rangle = \Lambda^2(t)+\frac{2K_\alpha}{\Gamma(\alpha)}\int_0^t d\tau
\frac{\tau^{\alpha-1}}{a^2(\tau)} &&\\
\label{x3WW}
\langle x^3\rangle =\Lambda^3(t)+\frac{6K_\alpha}{\Gamma(\alpha)}\Lambda(t)\int_0^t
d\tau\frac{\tau^{\alpha-1}}{a^2(\tau)} &&\\
\label{x4WW}
 \fl \langle x^4\rangle =\Lambda^4(t)+\frac{12K_\alpha}{\Gamma(\alpha)}\Lambda^2(t)\int_0
^td\tau\frac{\tau^{\alpha-1}}{a^2(\tau)}+24\frac{(K_{\alpha})^2}{\Gamma(\alpha)}
\int_0^t\frac{d\tau}{a^2(\tau)}\,_0D_{\tau}^{-\alpha}\frac{\tau^{\alpha-1}}{a^2(
\tau)}.&&
\end{eqnarray}
\end{subequations}
Focusing on the first two equations, equation \eqref{x1WW} tells us that the
first moment $\langle x\rangle$ is simply given by the deterministic shift,
i.e., the Lagrangian distance $\Lambda(t)$ travelled by ${\cal S}_0$ after
a time $t$. Conversely, equation \eqref{x2WW} implies that $\langle x^2\rangle_0
-\langle x\rangle_0^2$ equals $\langle x^2\rangle-\langle x\rangle^2$, i.e., the
variance in ${\cal S}_0$ is the same as in ${\cal S}_L$. Thus, regardless of the
value of $\alpha$ the dispersion properties of the random walk are not affected
by the velocity field. In contrast, in the case of a constant external force
\cite{levot1} this only holds when the particles are Brownian ($\alpha=1$).
As soon as $0<\alpha<1$, the dispersion properties are altered both on a static
domain \cite{report} and on a growing domain \cite{levot1}.

It is instructive to check that the expressions for the moments obtained above
can be directly recovered from equation \eqref{EcuWxExp}.
In order to deduce equations \eqref{x1WW} to \eqref{x3WW} from
equation \eqref{EcuWxExp}  we first multiply this latter equation with $x^n$ and
subsequently integrate over space. We are then left with the hierarchy
\begin{equation}
\fl
\frac{d\langle x^n\rangle}{dt}-n\nu\langle x^{n-1}\rangle=\frac{K_\alpha}{\Gamma(
\alpha)a^2(t)}\left(\frac{\partial}{\partial t}\int_0^t d\tau\frac{I^{(n)}_2}{(t-
\tau)^{1-\alpha}}+\nu\int_0^td\tau\frac{I^{(n)}_3}{(t-\tau)^{1-\alpha}}\right),
\end{equation}
where
\begin{equation}
I^{(n)}_m=\int_{-\infty}^\infty dxx^n\frac{\partial^m}{\partial x^m}W(x-\Lambda(t)
+\Lambda(\tau),\tau).
\end{equation}
Performing the change of variable $z=x-\Lambda(t)+\Lambda(\tau)$ in this integral, using the
binomial expansion for $z^n$, and integrating by parts, one finally obtains
\begin{equation}
\label{dxnEq}
\fl
\frac{d\langle x^n\rangle }{dt}=n\nu\langle x^{n-1}\rangle+\frac{K_\alpha}{
\Gamma(\alpha)a^2(t)}\sum_{j=2}^n {n\choose j} j(j-1)\left(\frac{d}{dt}
J^{(n)}_{j,2}-\nu(j-2)J^{(n)}_{j,3}\right),
\end{equation}
where
\begin{equation}
J^{(n)}_{j,m}=\int_0^t d\tau\frac{\left[\Lambda(t)-\Lambda(\tau)\right]^{n-j}}{
(t-\tau)^{1-\alpha}}\langle x^{j-m}(\tau)\rangle.
\end{equation}
For $n=1$ equation \eqref{dxnEq} becomes $d\langle x\rangle/dt=\nu$, and equation
\eqref{x1WW} follows immediately. For $n=2$ relation \eqref{dxnEq} becomes
\begin{equation}
\frac{d\langle x^2\rangle}{dt}=2\nu\langle x\rangle+\frac{2K_\alpha}{\Gamma(
\alpha)a^2(t)}\frac{d}{dt}J_{2,2}^{(2)}.
\end{equation}
From this equation and from result \eqref{x1WW} one immediately finds equation
\eqref{x2WW} by taking into account that $dJ_{2,2}^{(2)}/dt=t^\alpha/\alpha$.

We now proceed to derive equation \eqref{x3WW} from result \eqref{dxnEq}. For $n=3$,
relation \eqref{dxnEq} can be written as
\begin{equation}
\label{dx3dt2}
\frac{d\langle x^3\rangle}{dt}=3\nu\langle x^2\rangle+\frac{6K_\alpha}{\Gamma(
\alpha)a^2(t)}t^{\alpha-1}\Lambda(t)
\end{equation}
after a straightforward calculation making use of equation \eqref{x1WW}. Inserting
equation \eqref{x2WW} into \eqref{dx3dt2} one obtains
\begin{equation}
\label{dx3dt3}
\frac{d\langle x^3\rangle}{dt}=3\nu\Lambda^2+\frac{6K_\alpha}{\Gamma(\alpha)}
\frac{d}{dt}\left[\Lambda\int_0^t\frac{\tau^{\alpha-1}}{a^2(\tau)}d\tau\right].
\end{equation}
Equation \eqref{x3WW} then follows immediately from \eqref{dx3dt3}.

The validity of result \eqref{x4WW} for the fourth-order moment can also be
corroborated in a similar way. The calculation is straightforward but not
shown here explicitly.

\section{Mixing of diffusive pulses}
\label{sec:pulses}

We now proceed to study the influence of a velocity field on the mixing
properties of two diffusive pulses on a uniformly growing one-dimensional
domain. We consider both cases of normal and anomalous diffusion. For
convenience, our analysis will be carried out in Lagrangian coordinates.

As mentioned in the Introduction, mixing properties are crucial to understand
encounter-controlled reactions involving pairwise interactions, as originally
modelled by Smoluchowski \cite{smoluchowski}. In biological cells, for
instance, monomers of regulatory proteins meet diffusively and form dimers
\cite{ptashne}, and the dimer then diffuses to its designated binding site
on the genome or a DNA plasmid \cite{otto,aljaz}.\footnote{In the latter case
the diffusion coefficients of the two binding partners are different.} Analogous
processes need to run off in vesicles designed as artificial cells \cite{elani}.
In the case of particles advected by the growing domain, for instance, by the
expanding cytoskeleton in a living biological cell or compartments in growing
vesicles \cite{elani}, it is intuitively clear that the associated Hubble drift
will reduce diffusional mixing and thereby decrease the reaction rate
\cite{VEAY2018, EYAV2018, AEVY2019}. For sufficiently fast domain growth the lack of
mixing stems from a premature freezing of the pulse propagators
\cite{Yuste2016, VEAY2018, EYAV2018, AEVY2019},
since the Lagrangian step lengths become increasingly short in the course of time
(for a contracting domain, the Lagrangian steps become larger and one has the opposite effect).

In our problem, the random motion of each pulse is not only subject to a
Hubble drift, but also to an additional drift arising from the velocity
field. In order to formulate the problem in the most general form, we will
assume that each diffusing particle ``feels'' a different velocity field. In
other words, the force underlying this velocity field can be thought of as
being able to discriminate each particle by a distinctive property. For
instance, if the physical origin of the force is an electric field acting on
charged, overdamped diffusive particles, then this property will be the sign
and absolute value of their electric charge. In the particular case where both
particles have the same charge, they will experience the same biasing force,
and the mixing problem will be equivalent to its zero-field counterpart,
save for a coordinate shift. In the biological context, we could be thinking
of growing neuron cells, in which messenger RNA molecules are shuttled along
by molecular motors \cite{jaehyung}. Depending on the orientation of the
molecular track these motors are walking, their direction may be towards
either extremity of the pseudo-one-dimensional cell.

More specifically, consider two walkers labelled with indices $1$ and $2$. Let
$x_0^{(1)}$ and $x_0^{(2)}$ denote their initial positions. Without loss of
generality we assume that $x_0^{(1)}<x_0^{(2)}$. Owing to the external force
acting on each walker, the maxima are shifted with respect to their initial
positions as the pulses associated with each walker widen. Correspondingly,
one has $x_M^{(1)}=x_0^{(1)}+v^{(1)}T(t)$ and $x_M^{(2)}=x_0^{(2)}+v^{(2)}T(t)$.
The velocities $v^{(1)}$ and $v^{(2)}$ will hereafter be considered to be
constant for the sake of simplicity.

The two-pulse PDF can be written as a normalised linear combination of the
propagators for the respective initial conditions, i.e.,
\begin{equation}
\hspace{-1cm} W(x,t)=\frac{1}{2}\left[W_0^{(1)}(x-x_0^{(1)}-v^{(1)}T(t),t)+W_0^{(2)}(x-x_0^{(2)}
-v^{(2)}T(t),t)\right].
\label{WPulsesGen}
\end{equation}
Here, one encounters the difficulty that if at least one of the particles
is subdiffusive, its propagator is not known, and hence its contribution to
the joint PDF is also unknown. Nevertheless, it is possible to carry out a
semiquantitative study of particle mixing on the basis of the second-order
moments. To this end, let us first introduce the Lagrangian half-width of
a zero-field single-particle propagator as
\begin{equation}
\label{half-width-def}
w(t)=2\sqrt{\langle x^2\rangle_0-\langle x\rangle_0^2}.
\end{equation}
Let us further define two characteristic points $x_C^{(1)}(t)=x_M^{(1)}+w^{(1)}(t)$
and $x_C^{(2)}(t)=x_M^{(2)}-w^{(2)} (t)$ where, following the notation of equation
\eqref{half-width-def}, $w^{(1,2)}(t)>0$ denotes the Lagrangian half-width of the
symmetric propagator $W_0^{(1,2)}(x,t)$. Mixing after a time $t$ will be considered
to be weak (in a statistical sense) if $x_C^{(2)}(t)$ remains to the right of $x_C
^{(1)}(t)$, i.e., if the characteristic distance $d_C(t)\equiv x_C^{(2)}(t)-x_C^{
(1)}(t)$ remains $>0$. In other words, at a time $t$, one has weak mixing if
\begin{equation}
x_0^{(2)}-x_0^{(1)}>\left[v^{(1)}-v^{(2)}\right]T(t)+w^{(1)}(t)+w^{(2)}(t).
\label{WeakMixingCondition}
\end{equation}
In contrast, when
\begin{equation}
x_0^{(2)}-x_0^{(1)}<\left[v^{(1)}-v^{(2)}\right]T(t)+w^{(1)}(t)+w^{(2)}(t)
\label{MixingCondition}
\end{equation}
we will simply speak about ``mixing''. Similarly, if ``$<$'' can be replaced with
``$\ll$'' in equation \eqref{MixingCondition} we will speak about ``strong mixing''.
The last two terms on the right-hand side are positive and independent of the
respective velocities. Therefore, for a given set of diffusive properties, the
mixing behaviour depends on the first term, which is influenced by the relative
velocity $v^{(1)}-v^{(2)}$ and by the scale factor $a(t)$ entering the definition
of $T(t)$. Clearly, a positive (negative) relative velocity favours (hinders)
mixing, as is the case in the particular case of a static domain (this latter
case is recovered by setting $T(t)\equiv t$).

Note, however, that the domain growth introduces a key difference in the behaviour.
On a static domain the strong mixing condition is always fulfilled provided that
one waits long enough, irrespective of whether the diffusive pulses are normal or
anomalous (mixing becomes maximal when both pulse peaks overlap, i.e., when
the distance $d_M\equiv x_M^{(2)}-x_M^{(1)}$ between the two peaks vanishes).
However, for a given initial pulse separation and relative velocity, it is intuitively
clear that a sufficiently fast domain growth may
freeze both pulses before their mixing becomes significant. More precisely, for
$v^{(1)}>0$ and $v^{(2)}<0$, one has weak mixing at arbitrarily long times if
\begin{equation}
x_0^{(2)}-x_0^{(1)}>\left[v^{(1)}-v^{(2)}\right]T_\infty+w^{(1)}_\infty+w^{(2)}
_\infty,
\label{NonMixingCondition}
\end{equation}
where $T_\infty\equiv\lim_{t\to\infty} T(t)<\infty$ and $w^{(1,2)}_\infty=\lim_{
t\to\infty} w^{(1,2)}(t)<\infty$ are the long-time asymptotic values of the
half-widths. Thus, a sufficiently fast domain growth favours the localisation
of the Lagrangian propagators about the respective initial conditions, thereby
preventing that the memory of the latter is eventually lost by mixing.

In terms of physical coordinates, the situation described by equation
\eqref{NonMixingCondition} corresponds to the case when the Hubble
drift is so strong that the pulses separate from each other at a rate
much faster than the typical growth rate of their half-widths. Therefore,
the overlap of both pulses remains negligibly small at all times.

In what follows, we focus on the case of a set of two particles with identical
diffusive properties ($w\equiv w^{(1)}=w^{(2)}$) but subject to opposite drift
velocities $v\equiv v^{(1)}=-v^{(2)}$. Without loss of generality the midpoint
between the two pulse peaks will be chosen as the origin, i.e., $x_0\equiv x_0
^{(2)}=-x_0^{(1)}$.

\subsection{Normal diffusion and subdiffusion}

When both particles are normal-diffusive ($\alpha=1$) or subdiffusive ($0<
\alpha<1$) the variance $\langle x^2(t) \rangle_0$ is well-defined and can
be used to estimate $w(t)$.

In order to study the mixing kinetics the parameter $P$ in equation \eqref{tip-width-Levy}
should be chosen large enough to ensure significant mixing as soon as both tails overlap. In
general, for a given value of $P$ the corresponding value of $C_{\mu}$ (as well
as the associated characteristic width $w_{\mu}(t)=C_{\mu}\sigma_L(t)$) must be
computed numerically. However, for the Gaussian case $\mu=2$ one can obtain an
analytic expression, namely, $C_2=2\mathrm{erf}^{-1}(P)$. This result is in
agreement with what we had already anticipated for the Brownian case since, for
a Gaussian distribution, the probability that a particle is found within an
interval of half-width $w(t)=2\sigma(t)=2\sqrt{2}\sigma_L(t)$ is $P=0.9545$,
i.e., precisely the value which follows from the relation $\mathrm{erf}^{-1}(P)
=\sqrt{2}$.

In the case of two Gaussian pulses with identical diffusive properties and
opposite drift velocities this last result implies that when the weak mixing
condition \eqref{WeakMixingCondition} holds the overlap of both tails as
given by
\begin{equation}
\int_{-\infty}^{x_C^{(1)}}W_0^{(2)}(x,t)dx+\int_{x_C^{(2)}}^{\infty}W_0^{(1)}
(x,t)dx
\end{equation}
remains below $5\%$ at all times.

\begin{table}
\centering
\begin{tabular}{|c|c|c|c|c|c|}
\hline
& $\gamma<\alpha/2$ & $\gamma=\alpha/2$ & $\alpha/2<\gamma<1$ & $\gamma=1$ &
$\gamma>1$ \\
\hline\hline
$T$ & $t^{1-\gamma}$ & $t^{1-\gamma}$ & $t^{1-\gamma}$ & $\ln t$ & $\mathrm{
const.}$ \\\hline
$w$ & $t^{(\alpha-2\gamma)/2}$ & $\sqrt{\ln t}$ & $\mathrm{const.}$ & $\mathrm{
const.}$ & $\mathrm{const.}$ \\\hline
\end{tabular}
\caption{Asymptotic long-time behaviour of $T(t)$ and $w(t)$ for a power-law
scale factor with characteristic exponent $\gamma$. The results for $T(t)$
and $w(t)$ stem, respectively from the long-time behaviours of equations
\eqref{trans-time-pot} and \eqref{half-width-def}.}
\label{Table:Alpha}
\end{table}

For a power-law scale factor $a(t)\sim t^{\gamma}$ the different possible subcases
are given in Table \ref{Table:Alpha}. From this table one concludes that for $v=0$
and $x_0\ll w_\infty$ strong mixing does not occur when $\gamma>\alpha/2$. However,
a nonzero value of $v$ may completely change this scenario and bring about strong
mixing for sufficiently long times. For $v>0$, when $\alpha/2<\gamma\leq1$, one
has $T_\infty=\infty$, implying that the crossing of the two maxima will eventually
occur with certainty. In contrast, when the domain growth is fast enough to ensure
that $T_\infty<\infty$ the propagator evolves towards a steady state. In this latter
case strong mixing will never take place if it has not already occurred by the
characteristic time $t_C$ at which $W(x,t_C)$  can be considered to be practically
indistinguishable from $W(x,\infty)$. According to equation \eqref{MixingCondition}
strong mixing may be observed at a finite time $t$ if the initial separation
distance is small enough---more precisely, if $x_0\ll v T(t)+w(t)$ holds.

\begin{figure}
\centering
\includegraphics[width=8cm]{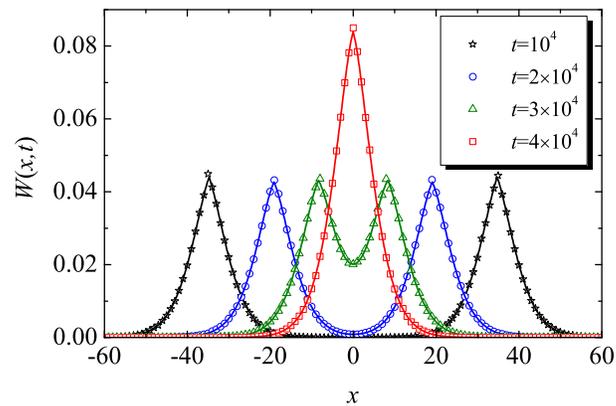}
\caption{Lagrangian propagator at times $t=10^4$, $2\times10^4$, $3\times10^4$,
and $4\times10^4$, for two subdiffusive pulses that drift in opposite directions.
Parameter values: $x_0 =75$, $\alpha=1/2$, and $K_{\alpha}=1/2$. The domain growth
is given by the power-law scale factor $a(t)=(1+t/t_0)^{\gamma}$ with $\gamma=3/4$
and $t_0=10^3$. The advective velocity of the pulses is $v= x_0/T_1(4\times10^4)
\approx1.225\times10^{-2}$. Solid lines: Theoretical curves obtained from the
numerical integration of the fractional diffusion equation. We used an adaptation of
the fractional Crank-Nicolson algorithm of reference \cite{yuste_weighted} with $\Delta t=0.1$
and $\Delta x=0.2$. The symbols depict simulations results ($10^6$ runs).}
\label{Fig:Mixing_t}
\end{figure}

Figure \ref{Fig:Mixing_t} displays the temporal evolution of two subdiffusive
pulses ($\alpha=1/2$) on a growing domain with scale factor $a(t)=(1+t/10^3)^{
3/4}$. Since $\gamma>\alpha/2$ we can see that the probability of overlap grows
in time. However, the respective pulse widths remain almost stationary throughout
the time window spanned by the represented set of plots (the length of this window
is $\sim10 t_0$). Thus, we conclude that in the present case mixing is driven by
the opposed velocity fields rather than by the spreading of the pulses. The maximum
overlap probability is attained after a time $t_M$, corresponding to the crossing
of both pulses, i.e., to a vanishing peak separation $d_{M}(t_M)=0$. This
characteristic time can be obtained from the solution of the implicit equation $x_0
=vT(t_M)$. Only at this specific time $t_M$ does the PDF correspond to the \emph{zero
field\/} solution for a single particle, $W(x,t_M)=W_0(x,t_M)$.

\begin{figure}
\centering
\includegraphics[width=8cm]{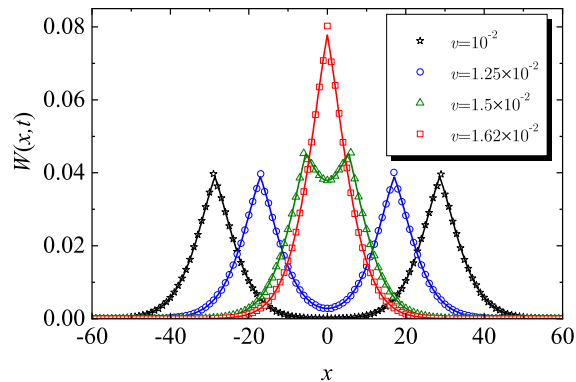}
\caption{Lagrangian propagator at times $t=10^4$ for $v=10^{-2}$, $1.25\times
10^{-2}$, $1.5\times10^{-2}$, and $v=x_0/T(10^4)\approx1.62\times10^{-2}$.
Solid lines: Propagator obtained from the numerical integration of the fractional
diffusion equation. We used a fractional Crank-Nicolson algorithm
with $\Delta t=0.1$ and $\Delta x=0.2$. The symbols depict simulations results ($10^6$ runs).
The remaining parameters are: $x_0 =75$, $\gamma=1/2$, $t_0=10^3$, $\alpha=1/2$, and $K_{\alpha}
=1/2$.}
\label{Fig:Mixing_v}
\end{figure}

For a fixed time $t=10^4$ figure \ref{Fig:Mixing_v} displays a series of snapshots
of the Lagrangian propagator corresponding to different values of $v$. For the
chosen value of the domain growth exponent ($\gamma=1/2$) and of the anomalous
diffusion exponent $\alpha$ the inequality $1/4=\alpha/2<\gamma<1$ holds, and so
the mixing is once again driven by the relative velocity term. As in the static
case, increasing $v$ results in enhanced mixing. For the chosen parameter set the
degree of mixing at a given time is maximised by the velocity value $v=0.162$. For
this precise value, $d_M(t=10^4)=0$, i.e., one has a single-hump propagator.

\begin{figure}
\centering
\includegraphics[width=8cm]{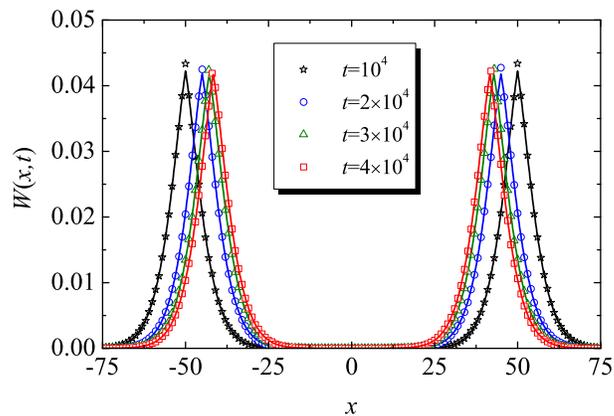}
\caption{Lagrangian propagator obtained from the numerical integration
of the fractional diffusion equation for $\alpha=1/2$, $K_{\alpha}=1/2$,
$a(t)=(1+t/5000)^2$, $x_0=75$, and $v=7.5\times10^{-3}$. We used a
fractional Crank-Nicolson algorithm with $\Delta t=0.1$ and $\Delta x=0.2$. The
symbols depict simulations results ($10^6$ runs).}
\label{Fig:NoMixing}
\end{figure}

Finally in figure \ref{Fig:NoMixing} we study a case where the domain growth is
so fast ($\gamma=2$) that mixing is absent at all times. As one can see, the
pulse widths remain practically the same at all times, and the deterministic
displacement of the pulse peaks is greatly slowed down by the domain growth,
until both pulses become almost stationary.

The theoretical curves in figures \ref{Fig:Mixing_t}-\ref{Fig:NoMixing} (obtained
by the numerical integration of the FDE) were corroborated by CTRW simulations.
As expected, a significant discrepancy
occurs at short times, since the number of steps taken by the random walker is not
large enough to reach the diffusive limit (see also reference \cite{report}). At longer
times the agreement is excellent.

\subsection{L\'evy flights}

To conclude our study about pulse mixing, let us now focus on the specific
case of L\'evy flights corresponding to the parameter choice $\alpha=1$
and $0<\mu<2$. The solution for a two-pulse initial condition can be easily
inferred from the one-particle propagator \eqref{Levyprop} for the zero-field
case via the relation \eqref{WPulsesGen}. The explicit form of the solution is
\begin{equation}
\fl
W(x,t)=\frac{1}{2}\left[\mathsf{L}_{\mu}\left(x+x_0-vT(t);[\sigma_L(t)]^{\mu}
\right)+\mathsf{L}_{\mu}\left(x-x_0+vT(t);[\sigma_L(t)]^{\mu}\right)\right],
\label{WEqPulsesLF}
\end{equation}
with $[\sigma_L(t)]^{\mu}=K_1^{\mu}\int_0^ta^{-\mu}(u)du$.

Let us once again consider the case of the power-law scale factor $a(t)=(1+t/t_0)
^\gamma$. As it turns out the asymptotic long-time behaviour of the typical width
depends on whether $1/\mu$ is larger, equal, or smaller than $\gamma$. Table
\ref{Table:Mu} summarises the typical subcases that result from the long-time
behaviour of $T(t)$ and $w_\mu(t)$ for different values of $\gamma$ and $\mu$.

\begin{table}
\centering
\begin{tabular}{|c|c|c|c|c|c|c|}
\hline
& $\gamma<\min\{1/\mu,1\}$ & $1/\mu=\gamma<1$ & $1/\mu<\gamma<1$ & $\gamma=1<1/\mu$
& $1/\mu=\gamma=1$ & $1/\mu<\gamma=1$\\\hline\hline
$T$ & $t^{1-\gamma}$ & $t^{1-\gamma}$ & $t^{1-\gamma}$ & $\ln t$ & $\ln t$ & $\ln
t$\\\hline
$w_\mu$ & $t^{1/\mu-\gamma}$ & $(\ln t)^{1/\mu}$ & $\mathrm{const.}$ & $t^{1/\mu-
\gamma}$ & $(\ln t)^{1/\mu}$ & $\mathrm{const.}$\\\hline
\end{tabular}
\begin{tabular}{|c|c|c|c|}
\hline
& $1<\gamma<1/\mu$ & $1<\gamma=1/\mu$ & $\gamma>\max\{1,1/\mu\}$\\\hline\hline
$T$ & $\mathrm{const.}$ & $\mathrm{const.}$ & $\mathrm{const.}$\\\hline
$w_{\mu}$ & $t^{1/\mu-\gamma}$ & $(\ln t)^{1/\mu}$ & $\mathrm{const.}$\\\hline
\end{tabular}
\caption{Asymptotic long-time behaviour of $T(t)$ and $w_{\mu}$ in the case of a
power-law expansion with exponent $\gamma$.}
\label{Table:Mu}
\end{table}

Within the permitted range $0<\mu<2$ it is convenient to distinguish two
subclasses of L\'evy flights with different qualitative behaviour, i.e.,
flights with $\mu\ge 1$, and flights with $\mu<1$. For the first subclass the
qualitative behaviour is similar to that of a Brownian process and is
essentially obtained by performing the replacement $1/2\to1/\mu$. Thus, for
$v\neq0$ the mixing of both pulses can be avoided for arbitrarily long times
by choosing $\gamma>1$. However, when $v=0$ one must only have
$\gamma>1/\mu$ in order to prevent mixing. The reason is that the
respective pulse widths grow as $t^{1/\mu}$, which is not fast enough to ensure significant
diffusive mixing of both pulses when their separation distance increases as $t^\gamma$ (with
$\gamma>1/\mu$) due to the domain growth.

In contrast, for L\'evy flights with $\mu<1$, the dominant contribution to
mixing in the long time limit will stem from L\'evy diffusion rather than from
the biasing fields. Note that such a regime can never occur when the jump length
PDF has a finite variance, i.e., in the normal-diffusive case or in the
subdiffusive case.

\begin{figure}
\centering
\includegraphics[width=8cm]{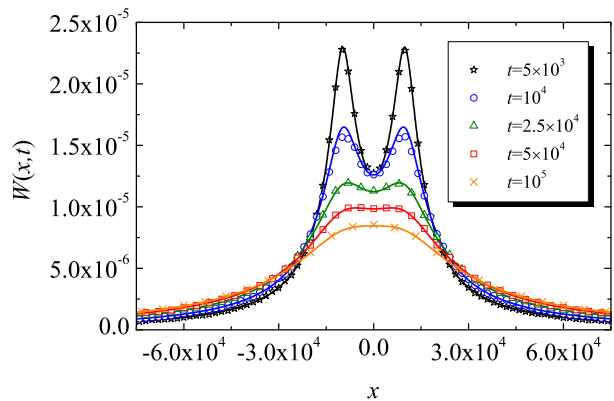}
\caption{Lagrangian propagator at times $t=5\times10^3$, $10^4$, $2.5\times10^4$,
$5\times10^4$, and $10^5$ for two L\'evy pulses with $\mu=3/4$, $K_1^{\mu}=1/2$,
$x_0=10^4$, and $v=1/10$ for a domain growth rate given by the scale factor $a(t)
=(1+t/10^3)^{5/4}$. Solid lines represent the theoretical solution obtained from
equation \eqref{WEqPulsesLF}. Symbols depict simulations results ($10^6$ runs).}
\label{Fig:Mixing_LF}
\end{figure}

The most clear-cut situation is found in the range $1<\gamma\leq1/\mu$. In this
regime, since $T_\infty<\infty$ the positions of both peaks tend to fixed
limiting values. The advection velocity $v$ and the initial pulse separation $2
x_0$ will determine whether or not such limiting values are attained before the
two peaks meet. In either case diffusive spreading remains dominant with respect
to the Hubble drift after a sufficiently long time and the pulse widths grow
without bound. Consequently, at long times the mixing process proceeds via the
widening of the respective pulses. Both pulses eventually merge into a single one,
which still continues to widen. This is precisely what is seen in figure
\ref{Fig:Mixing_LF}, which displays the evolution of the theoretical propagator
for two L\'evy pulses with characteristic exponent $\mu=3/4$ spreading on a
domain whose growth is controlled by a power-law factor with $\gamma=5/4$ and
$t_0=10^3$. The figure also shows simulations results that are in excellent
agreement with the theoretical expression \eqref{WEqPulsesLF}.

Finally, let us discuss the behaviour when $1<1/\mu<\gamma$. Here the diffusive
spreading remains the main mixing mechanism over a transient period. However,
the Hubble drift eventually becomes dominant, and consequently $W(x,t)$ tends to
a stationary (frozen) profile in the long-time limit. Depending on the
characteristic parameters ($\gamma$, $t_0$, $K_1^{\mu}$, $x_0$, and $v$) the
distribution $W(x,t)$  will be single- or double-peaked at given time.

\section{Summary and outlook}
\label{sec:summary}

We investigated several aspects of an anomalous diffusion processes described by
a separable CTRW evolving on a uniformly growing one-dimensional domain. First
we studied the behaviour of the first moments in the symmetric case, with special
focus on the rich phenomenology of the kurtosis. One of the most remarkable features
is that this quantity becomes time dependent in the subdiffusive case. As a
result of this, an initially non-Gaussian subdiffusive pulse was shown to exhibit
a Gaussian-like long-time behaviour for a suitable parameter choice.

We subsequently considered the effect of a velocity field on this CTRW dynamics. We
derived the corresponding FDAE \eqref{EcuWxExp}, which generalises a previous
result valid for the special case $\mu=2$ on a static domain\cite{Cairoli2018}.
Indeed, since our bifractional equation holds also for $\mu<2$, it includes L\'evy
flights as a particular case. The mathematical form of the FDAE is rather peculiar
and not intuitive even in the case of a static domain, given the very straightforward
Galilean transformation \eqref{bias-no-bias} between the walker's PDF in the lab
frame $\mathcal{S}_L$ and its counterpart in the comoving frame $\mathcal{S}_0$.

Taking the above results as a starting point we studied the mixing behaviour of a
pair of diffusive pulses in a one-dimensional domain whose time evolution is
governed by a power-law scale factor. We focused on the case where the pulses are
drifting with velocities $v$ and $-v$ as they spread, the spreading of each pulse
being characterised by their respective half-widths. In this scenario a sufficiently
fast domain growth was found to largely prevent mixing between a pair of normal
diffusive walkers or between a pair of subdiffusive walkers. However, a sufficiently
large value of $v$ is able to restore mixing. In the superdiffusive case, the
behaviour is more complex. For $\mu\ge 1$ and a sufficiently large initial separation
of the pulses mixing is essentially controlled by the relative velocity introduced
by the fields, as is the case for normal diffusive or for subdiffusive walkers. In
contrast, when $\mu<1$, diffusional mixing dominates over the deterministic mixing
induced by the velocity fields.

The present work may be extended in several directions. One such direction should
consider the mixing behaviour of pulses for time dependent velocity fields $v=v(t)$.
Another interesting generalisation concerns the case of non-separable CTRWs, e.g.,
L\'evy walks. Finally, one could explore the effect of nonuniform domain growth
\cite{abad} introducing a spatial dependence in the scale factor.

\ack

E.~A., F.~L.~V., and S.~B.~Y. acknowledge support by the Spanish Agencia Estatal
de Investigaci\'on Grant (partially financed by the ERDF) No.~FIS2016-76359-P
and by the Junta de Extremadura (Spain) Grant (also partially financed by the
ERDF) No.~GR18079. In addition, F.~L.~V. acknowledges financial support from
the Junta de Extremadura through Grant No. PD16010 (partially financed by FSE
funds). R.~M. acknowledges the German Research Foundation (DFG) grant number
ME 1535/7-1 as well as the Foundation for Polish Science (Fundacja na rzecz
Nauki Polskiej, FNP) for an Alexander von Humboldt Polish Honorary Research
Scholarship.


\section*{References}

\begin{thebibliography}{99}

\bibitem{montroll} E. W. Montroll and G. H. Weiss, \emph{Random Walks on Lattices. III. Calculation of First-Passage Times with Application to Exciton Trapping on Photosynthetic Units}, J. Math. Phys. \textbf{10}, 753 (1969).

\bibitem{scher} H. Scher and E. W. Montroll, \emph{Anomalous transit-time dispersion in amorphous solids}, Phys. Rev. B \textbf{12}, 2455 (1975).

\bibitem{scher1} H. Scher  and M. Lax, \emph{Stochastic Transport in a Disordered Solid. I. Theory}, Phys. Rev. B \textbf{7}, 4491 (1973).

\bibitem{henning} H. Kr{\"u}semann, R. Schwarzl, and R. Metzler, \emph{Ageing
Scher-Montroll transport}, Transp. Porous Media \textbf{115}, 327 (2016).

\bibitem{barzykin} A. V. Barzykin and M. Tachiya, \emph{Luminiscence Quenching in Micellar Clusters
as a Random Walk Problem}, Phys. Rev. Lett. \textbf{73}, 3479 (1994).

\bibitem{hornung} G. Hornung, B. Berkowitz, and N. Barkai, \emph{Morphogen gradient formation in a
complex environment: An anomalous diffusion model}, Phys. Rev. E \textbf{72}, 041916 (2005).

\bibitem{yuste} S. B. Yuste, E. Abad, and K. Lindenberg, \emph{Reaction-subdiffusion model of morphogen
gradient formation}, Phys. Rev. E \textbf{82}, 061123 (2010).

\bibitem{ling} P. Tan, Y. Liang, Q. Xu, E. Mamontov, J. Li, X. Xing, and L. Hong,
\emph{Gradual crossover from subdiffusion to normal diffusion: a many-body effect
in protein surface water}, Phys. Rev. Lett. \textbf{120}, 248101 (2018).

\bibitem{smith} X. Hu, L. Hong, M. D. Smith, T. Neusius, X. Cheng, and J. C.
Smith, \emph{The Dynamics of Single Protein Molecules is Non-Equilibrium and
Self-Similar over Thirteen Decades in Time}, Nature Phys. \textbf{12}, 171 (2015).

\bibitem{diego} A. V. Weigel, B. Simon, M. M. Tamkun, and D. Krapf, \emph{Ergodic
and Nonergodic Processes Coexist in the Plasma Membrane as Observed by
Single-Molecule Tracking}, Proc. Natl. Acad. Sci. U.S.A. \textbf{108}, 6438 (2011).

\bibitem{stas} S. M. A. Tabei, S. Burov, H. Y. Kim, A. Kuznetsov, T. Huynh,
J. Jureller, L. H. Philipson, A. R. Dinner, and N. F. Scherer, \emph{Intracellular
Transport of Insulin Granules is a Subordinated Random Walk}, Proc. Natl. Acad. Sci.
U.S.A. \textbf{110}, 4911 (2013).

\bibitem{lene} J.-H. Jeon, V. Tejedor, S. Burov, E. Barkai, C. Selhuber-Unkel, K.
Berg-S{\o}rensen, L. Oddershede, and R. Metzler, \emph{In Vivo Anomalous Diffusion
and Weak Ergodicity Breaking of Lipid Granules}, Phys. Rev. Lett. \textbf{106},
048103 (2011).

\bibitem{granick} K. Chen, B. Wang, and S. Granick, \emph{Memoryless self-reinforcing
directionality in endosomal active transport within living cells}, Nature Mat.
\textbf{14}, 589 (2015).

\bibitem{jae} M. S. Song, H. C. Moon, J.-H. Jeon, and H. Y Park, \emph{Neuronal
messenger ribonucleoprotein transport follows an aging L{\'e}vy walk}, Nature
Comm. \textbf{9}, 344 (2018).

\bibitem{edely} Y. Edery, A. Guadagnini, H. Scher, and B. Berkowitz,
\emph{Origins of anomalous transport in heterogeneous media:
Structural and dynamic controls}, Wat. Resour. Res. \textbf{50}, 1490 (2014).

\bibitem{grl} H. Scher, G. Margolin, R. Metzler, J. Klafter, and B. Berkowitz,
\emph{The dynamical foundation of fractal stream chemistry: The origin
of extremely long retention times}, Geophys. Res. Lett. \textbf{29}, 1061 (2002).

\bibitem{scalas} E. Scalas, R. Gorenflo, and F. Mainardi, \emph{Fractional calculus and
continuous time finance}, Physica A \textbf{284}, 376 (2000).

\bibitem{masoliver} J. Masoliver, M. Montero, J. Perell\'o, and G. H. Weiss,
\emph{The continuous time random walk formalism in financial markets}, J. Econ. Behav. Organ.
\textbf{61}, 577 (2006).

\bibitem{pt-scher} H. Scher, M. F. Shlesinger, and J. T. Bendler, \emph{Time-scale
invariance in transport and relaxation}, Phys. Today \textbf{44}, 26 (1991).

\bibitem{pt-yossi} J. Klafter, M. F. Shlesinger, and G. Zumofen, \emph{Beyond
Brownian motion}, Phys. Today \textbf{49}, 33 (1996).

\bibitem{nature-mike} M. F. Shlesinger, G. M. Zaslavsky, and J. Klafter,
\emph{Strange kinetics}, Nature \textbf{363}, 31 (1993).

\bibitem{bouchaud} J.-P. Bouchaud and A. Georges, \emph{Anomalous diffusion in
disordered media - statistical mechanisms, models and physical applications},
Phys. Rep. \textbf{195}, 127 (1990).

\bibitem{report} R. Metzler and J. Klafter, \emph{The random walk's guide to anomalous
diffusion: a fractional dynamics approach}, Phys. Rep. \textbf{339}, 1 (2000).

\bibitem{report1} R. Metzler and J. Klafter, \emph{The restaurant at the end of the
random walk: recent developments in the description of anomalous transport by
fractional dynamics}, J. Phys. A \textbf{37}, R161 (2004).

\bibitem{igorsm} I. M. Sokolov, \emph{Models of anomalous diffusion in crowded
environments}, Soft Matter \textbf{8}, 9043 (2012).

\bibitem{pccp} R. Metzler, J.-H. Jeon, A. G. Cherstvy, and E. Barkai, \emph{Anomalous
diffusion models and their properties: non-stationarity, non-ergodicity, and
ageing at the centenary of single particle tracking}, Phys. Chem.  Chem. Phys.
\textbf{16}, 24128 (2014).

\bibitem{pt1} E. Barkai, Y. Garini, and R. Metzler, \emph{Strange Kinetics of Single
Molecules in Living Cells}, Phys. Today \textbf{65}, 29 (2012).

\bibitem{pt2} D. Krapf and R. Metzler, \emph{Strange interfacial molecular dynamics},
Phys. Today \textbf{72}, 48 (2019).

\bibitem{hughes} B. D. Hughes, \emph{Random walks and random environments}, vol.~1
(Oxford University Press, Oxford UK, 1990).

\bibitem{bouchaud_web} J.-P. Bouchaud, \emph{Weak ergodicity breaking and
aging in disordered systems}, J. Phys. (Paris) I \textbf{2}, 1705 (1992).

\bibitem{eli_bel} G. Bel and E. Barkai,
\emph{Weak Ergodicity Breaking in the Continuous-Time Random Walk},
Phys. Rev. Lett. \textbf{94}, 240602 (2005).

\bibitem{eli_rebenstok} A. Rebenshtok and E. Barkai, \emph{Distribution of Time-Averaged Observables for Weak Ergodicity Breaking}, Phys. Rev. Lett. \textbf{99}, 210601 (2007).

\bibitem{irwin} M. A. Lomholt, I. M. Zaid, and R. Metzler, \emph{Subdiffusion and Weak Ergodicity Breaking in the Presence of a Reactive Boundary}, Phys. Rev. Lett. \textbf{98}, 200603 (2007).

\bibitem{stas_pnas} S. Burov, R. Metzler, and E. Barkai, \emph{Aging and non-ergodicity
beyond the Khinchin theorem}, Proc. Natl. Acad. Sci. U.S.A. \textbf{107}, 13228 (2010).

\bibitem{bouchaud_age} C. Monthus and J.-P. Bouchaud, \emph{Models of traps and glass phenomenology}, J. Phys. A \textbf{29}, 3847 (1996).

\bibitem{eli_age} E. Barkai, \emph{Aging in Subdiffusion Generated by a Deterministic Dynamical System}, Phys. Rev. Lett. \textbf{90}, 104101 (2003).

\bibitem{johannes} J. H. P. Schulz, E. Barkai, and R. Metzler,
\emph{Aging Effects and Population Splitting in Single-Particle Trajectory Averages}, Phys. Rev. Lett.
\textbf{110}, 020602 (2013).

\bibitem{johannes1} J. H. P. Schulz, E. Barkai, and R. Metzler,
\emph{Aging Renewal Theory and Application to Random Walks}, Phys. Rev. X
\textbf{4}, 011028 (2014).

\bibitem{manzo} C. Manzo, J. A. Torreno-Pina, P. Massignan, G. J. Lapeyre, Jr.,
M. Lewenstein, and M. F. Garcia Parajo,
\emph{Weak Ergodicity Breaking of Receptor Motion in Living Cells Stemming
from Random Diffusivity}, Phys. Rev. X \textbf{5}, 011021 (2015).

\bibitem{yossi_prl} J. Klafter and R. Silbey, \emph{Derivation of the Continuous-Time
Random-Walk Equation}, Phys. Rev. Lett. \textbf{44}, 55 (1980).

\bibitem{compte} A. Compte, \emph{Stochastic foundations of fractional dynamics},
Phys. Rev. E \textbf{53}, 4191 (1996).

\bibitem{mebakla} R. Metzler, E. Barkai, and J. Klafter,
\emph{Anomalous diffusion and relaxation close to thermal equilibrium:
A fractional Fokker-Planck equation approach},
Phys. Rev. Lett. \textbf{82}, 3563 (1999).

\bibitem{mebakla1} R. Metzler, E. Barkai, and J. Klafter,
\emph{Deriving fractional Fokker-Planck equations from a
generalised master equation}, Europhys. Lett. \textbf{46}, 431 (1999).

\bibitem{meklabvp} R. Metzler and J. Klafter,
\emph{Boundary value problems for fractional diffusion equations},
Physica A \textbf{278}, 107 (2000).

\bibitem{katja} E. Abad, S. B. Yuste, and K. Lindenberg, \emph{Reaction-subdiffusion and reaction-superdiffusion equations for evanescent particles performing continuous-time random walks},
    Phys. Rev. E \textbf{81}, 031115 (2010).

\bibitem{mortal} S. B. Yuste, E. Abad, and K. Lindenberg,
\emph{Exploration and Trapping of Mortal Random Walkers}, Phys. Rev. Lett. \textbf{110}, 220603 (2013).

\bibitem{evanescent} E. Abad, S. B. Yuste, and K. Lindenberg,
\emph{Evanescent continuous time random walks}, Phys. Rev. E \textbf{88}, 062110 (2013).

\bibitem{yossi_ws} J. Klafter, S.-C. Lim, and R. Metzler, Editors, \emph{Fractional
Dynamics in Physics} (World Scientific, Singapore, 2011).

\bibitem{henry} B. I. Henry and S. L. Wearne, \emph{Fractional reaction-diffusion}, Physica A \textbf{276}, 448 (2000).

\bibitem{igor_francesc} I. M. Sokolov, M. G. W. Schmidt, and F. Sagu\'es, \emph{Reaction-subdiffusion equations}, Phys. Rev. E \textbf{73}, 031102 (2006).

\bibitem{henry1} B. I. Henry, T. A. M. Langlands, and S. L. Wearne, \emph{Anomalous diffusion with linear reaction dynamics: From continuous time random walks to fractional reaction-diffusion equations}, Phys. Rev. E \textbf{74}, 031116 (2006).

\bibitem{yadav} A. Yadav and W. Horsthemke, \emph{Kinetic equations for reaction-subdiffusion systems: Derivation and stability analysis}, Phys. Rev. E \textbf{74}, 066118 (2006).

\bibitem{seki} K. Seki, A. I. Shushin, M. Wojcik, and M. Tachiya, \emph{Specific features of the kinetics of fractional-diffusion assisted geminate reactions}, J. Phys. Condens. Matter,
\textbf{19}, 065117 (2007).

\bibitem{fedotov} S. Fedotov and A. Iomin, \emph{Probabilistic approach to a proliferation and migration dichotomy in tumor cell invasion}, Phys. Rev. E \textbf{77}, 031911 (2008).

\bibitem{yadav1} A. Yadav, S. M. Milu, and W. Horsthemke, \emph{Turing instability in reaction-subdiffusion systems}, Phys. Rev. E \textbf{78}, 026116 (2008).

\bibitem{gafi} V. Gafiychuk, B. Datsko, and V. Meleshko, \emph{Mathematical modeling of time fractional reaction-diffusion systems}, J. Comput. Appl. Math. \textbf{220}, 215 (2008).

\bibitem{froemberg} D. Froemberg and I. M. Sokolov, \emph{Stationary fronts in an A + B --> 0 reaction under subdiffusion}, Phys. Rev. Lett. \textbf{100}, 108304 (2008).

\bibitem{MendezBook} V. M\'endez, S. Fedotov, and W. Horsthemke, \emph{Reaction-Transport Systems:
Mesoscopic Foundations, Fronts, and Spatial Instabilities} (Springer-Verlag, 2010).

\bibitem{levot} F. Le Vot, E. Abad, and S. B. Yuste, \emph{Continuous-time random-walk
model for anomalous diffusion in expanding media}, Phys. Rev. E \textbf{96},  032117 (2017).

\bibitem{angst} C. N. Angstmann, B. I. Henry, and A. V. McGann, \emph{Generalized fractional diffusion
  equations for subdiffusion in arbitrarily growing domains}, Phys. Rev. E \textbf{96}, 042153 (2017).

\bibitem{levot1} F. Le Vot and S. B. Yuste, Continuous-time random walks and Fokker-Planck equation in expanding media, Phys. Rev. E. \textbf{98}, 042117 (2018).

\bibitem{abad} E. Abad, C. N. Angstmann, B. I. Henry, A. V. McGann, F. Le Vot, and S. B. Yuste, \emph{Reaction-diffusion and reaction-subdiffusion equations on arbitrarily evolving domains},
 	arXiv:2002.06011.

\bibitem{Yuste2016} S.B. Yuste, E. Abad, and C. Escudero, \emph{Diffusion in an expanding medium: Fokker-Planck equation, Green's function, and first-passage properties}, Phys. Rev. E \textbf{94}, 032118 (2016).

\bibitem{VEAY2018}
F. Le Vot, C. Escudero, E. Abad, and S. B. Yuste,
\emph{Encounter-controlled coalescence and annihilation on a one-dimensional growing domain},
Phys. Rev. E \textbf{98} 032137 (2018).

\bibitem{EYAV2018}
C. Escudero, S. B. Yuste, E. Abad, and F. Le Vot, \emph{Reaction-diffusion kinetics in
  growing domains}, Handbook of Statistics \textbf{39}, 131 (2018).

\bibitem{AEVY2019}
E. Abad, C. Escudero,  F. Le Vot, and S. B. Yuste, First-Passage Processes and Encounter-Controlled Reactions in Growing Domains, in \emph{Chemical Kinetics: Beyond the Textbook}, edited by K. Lindenberg, R. Metzler, and G. Oshanin, pp. 409-433 (World Scientific, Singapore, 2019).

\bibitem{alberts} B. Alberts, A. Johnson, J. Lewis, D. Morgan, M. Raff, K. Roberts, and P. Walter, \emph{Molecular biology of the cell}, 6th ed. (Garland Science, New York, 2015).

\bibitem{biofilm} C. Wilson et al, \emph{Quantitative and Qualitative Assessment Methods
for Biofilm Growth: A Mini-review}, Res. Rev. J. Eng. Technol. \textbf{6}, 1 (2017).

\bibitem{biofilm1} H.-C. Flemming et al, \emph{Biofilms: an emergent form of bacterial life},
Nature Rev. Microbiol. \textbf{14}, 563 (2016).

\bibitem{tissue} D. Ambrosi, M. Ben Amar, C. J. Cyron, A. DeSimone, A. Goriely, J. D. Humphrey and E. Kuhl, \emph{Growth and remodelling of living tissues: perspectives, challenges and opportunities}, J. R. Soc. Interface \textbf{16},
20190233 (2019).

\bibitem{tissue1} S. C. Cowin, \emph{Tissue growth and remodeling}, Annu. Rev. Biomed. Eng.
\textbf{6}, 77 (2004).

\bibitem{vesicle} J. W. Szostak, D. P. Bartel, and P. L. Luisi, \emph{Synthesizing life},
Nature \textbf{409}, 387 (2001).

\bibitem{vesicle1} C. Xu, S. Hu, and X. Chen, \emph{Artificial cells: from basic science to
applications}, Mater. Today \textbf{19}, 516 (2016).

\bibitem{vesicle2} B. L.-S. Mui, P. R. Cullis, E. Evans, and T. D. Madden,
\emph{Osmotic properties of large unilamellar vesicles prepared by extrusion},
Biophys. J. \textbf{64}, 443 (1993).

\bibitem{carrier} O. Carrier, N. Shahidzadeh-Bonn, R. Zargar, M. Aytouna, M.
Habibi, J. Eggers, and D. Bonn, \emph{Evaporation of water: evaporation rate and
collective effects}, J. Fluid Mech. \textbf{798}, 774 (2016).

\bibitem{yang} Q. Yang and B. R. Scanion, \emph{How much water can be captured from
flood flows to store in depleted aquifers for mitigating floods and droughts?
A case study from Texas, US}, Environ. Res. Lett. \textbf{14}, 054011 (2019).

\bibitem{zhang} G. Zhang, G. Feng, X. Li, C. Xie, and X. Pi, \emph{Flood Effect on
Groundwater Recharge on a Typical Silt Loam Soil}, Water \textbf{9}, 523 (2017).

\bibitem{allegre}	C. J. Allègre and D. L. Turcotte, \emph{Implications of a two-component marble-cake mantle}, Nature \textbf{323}, 123 (1986).

\bibitem{kellogg} L. H. Kellogg and D. L. Turcotte, \emph{Homogenization of the mantle by convective mixing and diffusion}, Earth Planet. Sci. Lett. \textbf{81}, 371 (1987).

\bibitem{berezinsky} V. Berezinsky and A. Z. Gazizov, \emph{Diffusion of Cosmic Rays in the Expanding Universe. II. Energy Spectra of Ultra-High Energy Cosmic Rays}, Astrophys. J.  \textbf{669}, 684 (2007).

\bibitem{batista}  R. A. Batista and G. Sigl, \emph{Diffusion of cosmic rays at EeV energies in inhomogeneous extragalactic magnetic fields}, J. Cosmol. Astropart. Phys. \textbf{1},  031 (2014).

\bibitem{LVYA2019} F. Le Vot, E. Abad, and S. B. Yuste,
\emph{Standard and fractional Ornstein-Uhlenbeck process on a growing domain}, Phys. Rev. E \textbf{100}, 012142 (2019).

\bibitem{jaehyung} M. S. Song, H. C. Moon, J.-H. Jeon, and H. Y. Park,
\emph{Neuronal messenger ribonucleoprotein transport follows an aging L{\'e}vy walk},
Nature Comm. \textbf{9}, 344 (2018).

\bibitem{chen} Y. Chen, X. Wang, and W. Deng,  \emph{Subdiffusion in an external force field},  Phys. Rev. E \textbf{99}, 042125 (2019).

\bibitem{Cairoli2018} A. Cairoli, R. Klages, and A. Baule, \emph{Weak Galilean invariance as a selection principle for coarse-grained diffusive models}, Proc. Natl. Acad. Sci. U.S.A. \textbf{115}, 5714 (2018).

\bibitem{Podlubny1999}  I. Podlubny, \emph{Fractional Differential Equations: An Introduction to Fractional Derivatives, Fractional Differential Equations, to Methods of Their Solution and Some of Their Applications} (Academic Press, San Diego, 1999).

\bibitem{mathai} A. M. Mathai, R. K. Saxena, and H. J. Haubold, \emph{The H-Function:
Theory and Applications} (Springer, Berlin, 2009).

\bibitem{Abra72} M. Abramowitz and I. Stegun, \emph{Handbook of Mathematical Functions} (Dover publications, New York, 1972).

\bibitem{yuste_weighted} S. B. Yuste, \emph{Weighted average finite difference methods for fractional diffusion equations}, J. Comput. Phys. \textbf{216}, 264 (2006).

\bibitem{Sokolov2003} I. M. Sokolov and R. Metzler, \emph{Towards deterministic equations for L\'evy walks: The fractional material derivative}, Phys. Rev. E \textbf{98}, 042117 (2003).

\bibitem{Uchaikin2014} V. V. Uchaikin, R. T. Sibatov, and A. N. Byzykchi, \emph{Cosmic rays propagation along solar magnetic field lines: a fractional approach}, Commun. Appl. Ind. Math. \textbf{6},  e-479 (2014).

\bibitem{Metzler1998} R. Metzler, J. Klafter, and I. M. Sokolov, \emph{Anomalous transport in external fields: Continuous time random walks and fractional diffusion equations extended}, Phys. Rev. E \textbf{58}, 1621 (1998).

\bibitem{Fa2011} K. S. Fa,\emph{ Continuous time random walk: Galilei invariance and
relation for the \emph{n}th moment}, J. Phys. A: Math. Theor. \textbf{44}, 035004 (2011).

\bibitem{Ryden2003} B. Ryden, \textit{Introduction to Cosmology} (Addison-Wesley, Reading, 2003).

\bibitem{smoluchowski} M. v. Smoluchowski, \emph{Three presentations on diffusion,
molecular movement according to Brown and coagulation of colloid particles},
Physikal. Zeitschr. \textbf{17}, 557 (1916).

\bibitem{ptashne} M. Ptashne, \emph{A Genetic Switch, Phage Lambda Revisited}
(Cold Spring Harbor Laboratory Press, Cold Spring Harbor, 2004).

\bibitem{otto} O. Pulkkinen and R. Metzler, \emph{Distance matters: the impact of
gene proximity in bacterial gene regulation}, Phys. Rev. Lett.  \textbf{110},
198101 (2013).

\bibitem{aljaz} A. Godec and R. Metzler, \emph{Universal proximity effect in target
search kinetics in the few encounter limit}, Phys. Rev. X \textbf{6}, 041037 (2016).

\bibitem{elani} Y. Elani, R. V. Law, and O. Ces, \emph{Protein synthesis in artificial
cells: using compartmentalisation for spatial organisation in vesicle bioreactors},
Phys. Chem. Chem. Phys. \textbf{17}, 15534 (2015).

\end{thebibliography}
\end{document}